\documentclass[twocolumn]{aastex631}

\usepackage[english]{babel}
\usepackage[utf8]{inputenc}

\usepackage{amssymb}
\usepackage{hyperref}
\usepackage{lipsum}
\usepackage{amsmath}
\usepackage{graphicx}
\usepackage{physics}

\usepackage{natbib}

\newcommand{\REV}[1]{\textcolor{black}{#1}}

\shorttitle{The PgLis model}
\shortauthors{Pelosi et al.}

\graphicspath{{jasr}}

\begin{document}

\title{\textbf{A new model for long-term forecasting of Galactic cosmic rays}
}

\correspondingauthor{David Pelosi}
\email{david.pelosi@dottorandi.unipg.it}

\author[0009-0003-4663-1262]{David Pelosi} 
\affiliation{Università degli Studi di Perugia, Perugia 06100, Italy}
\affiliation{INFN - Perugia, Perugia 06100, Italy}

\author{Fernando Barão} 
\affiliation{Laboratório de Instrumentação e Física Experimental de Partículas, Lisboa 1000, Portugal}

\author{Bruna Bertucci} 
\affiliation{Università degli Studi di Perugia, Perugia 06100, Italy}
\affiliation{INFN - Perugia, Perugia 06100, Italy}

\author{Emanuele Fiandrini} 
\affiliation{Università degli Studi di Perugia, Perugia 06100, Italy}
\affiliation{INFN - Perugia, Perugia 06100, Italy}

\author[0000-0003-1874-2144]{Miguel Orcinha} 
\affiliation{INFN - Perugia, Perugia 06100, Italy}


\author{Nicola Tomassetti} 
\affiliation{Università degli Studi di Perugia, Perugia 06100, Italy}
\affiliation{INFN - Perugia, Perugia 06100, Italy}

\date{\today}

\begin{abstract}
\REV{The modulation of galactic cosmic rays, driven by the evolution of the heliospheric magnetic field, strongly influences the intensity of cosmic rays reaching near-Earth space. Characterizing this process is crucial both for advancing our understanding of cosmic-ray transport and for assessing radiation exposure and related hazards in space environments. 
Here we present a newly developed forecasting framework built on a numerical description of charged particle transport in the heliosphere and its dependence on solar activity, designed for the long-term forecasting of galactic cosmic-ray fluxes. It solves a one-dimensional, spherically symmetric form of the Parker transport equation, including diffusion, solar-wind advection, and adiabatic energy losses. The model has been validated using multi-species flux measurements from space-based experiments: PAMELA, AMS-02, and ACE. Its strategy is based on Hilbert-Huang transform filtering and cross-correlation between delayed solar proxies and effective model parameters. 
Our charge-sign- and rigidity-dependent parametric description of the diffusion-advection processes yields good overall agreement with the data, as shown by the reconstruction uncertainty.
The robustness of this approach is validated across a broad set of multichannel datasets covering different particle species, energy ranges, and phases of solar activity, supporting its applicability to space radiation monitoring and forecasting.
Furthermore, when coupled with solar-proxy forecasting models, it enables decadal-scale predictions of galactic cosmic-ray fluxes, thereby supporting long-term planning and radiation-risk assessment for future space missions.}
    
\end{abstract}



\section{Introduction}
Galactic cosmic rays (GCRs) are highly energetic charged particles originating from a variety of astrophysical sources within our Galaxy.
\REV{As they propagate through the heliosphere, they are exposed to the turbulent interplanetary magnetic field (IMF), which affects their trajectories and energy, as they are diffusely transported through it. 
This magnetic field, originating from the Sun, is carried outward by the solar wind, a supersonic stream of highly conductive plasma, which covers the entire solar system.
Among the effects that GCR endure are spatial diffusion, advection in the outward expanding solar wind, magnetic gradient and curvature drifts, and adiabatic energy losses, altering GCR intensity and reshaping their energy spectrum. This produces a net reduction of the GCR intensity reaching the inner heliosphere in a process known as \emph{solar modulation}. It inherits the temporal variability of solar activity, following the 11-year solar cycle, proxied by solar observables such as the sunspot number (SSN).}

Enhanced magnetic turbulence during the solar activity maxima strengthens modulation and reduces the near-Earth fluxes, while solar minima correspond to weaker solar shielding, highlighting a well-known anti-correlation pattern \cite{SOLARMOD_Potgieter_review}.
Understanding this variability is crucial for interpreting GCR origin, predicting space radiation environments, and assessing exposure risks for spacecraft and crew \cite{space_radiation_Dobney_2023}.
A key aspect of modulation studies is the time lag between solar activity and GCR intensity. Observations show delays of several months between SSN variations and neutron monitor rates \cite{Lag_General}, reflecting the time required for heliospheric plasma and magnetic structures to propagate and influence cosmic-ray transport. This lag is commonly attributed to a complex interplay of transport processes and evolving heliospheric conditions \cite{Wang_Lag}.

\subsection{Cosmic-ray datasets}
A joint effort combining long-term neutron monitor observations, stratospheric balloon measurements, and space-borne experiments has been essential for advancing our understanding of solar modulation and its temporal evolution. In addition, radiation-dose detectors operating in low-Earth orbit and outside the magnetosphere provide complementary datasets that enable valuable cross-checks against model predictions \cite{LIDAL_Romoli_2023, CRaTER_deWet_2020}. For the evaluation of the Local Interstellar Spectrum (LIS) of GCRs before entering the heliosphere, a key input for all solar modulation models, deep-space probes such as Voyager 1 and Voyager 2 have played a crucial role in constraining and validating flux models for a wide range of cosmic-ray nuclei \cite{Cummings_2016, Cummings_2025}. In this work, we rely on space-borne datasets for both training and testing. These include measurements from the Electron Proton Helium Instrument (EPHIN) onboard SOHO \cite{SOHO}, balloon-borne detectors such as BESS \cite{BESS97, BESS_04_07}, and satellite missions like PAMELA \cite{PAMELA_first}. We also incorporate continuous observations from the Cosmic Ray Isotope Spectrometer (CRIS) on ACE \cite{ACE/CRIS_first} and high-precision multichannel data from AMS-02 \cite{AMS02_first}, which provide daily and monthly fluxes for multiple species. We further leverage the temporal overlap among these datasets to extend both the energy range and species coverage, enhancing the robustness of the calibration and enabling more accurate modeling of cosmic-ray transport.


\subsection{GCR modulation models}
\REV{
GCR models have been developed both as empirical formulations and as physics-based solutions of the cosmic-ray transport equation, with varying levels of complexity and parameters involved. Among the latter are physical approximations such as the Force-Field and Convective-Diffusion approximations \cite{NTPRL, cholis_2016, zhu_2025, corti_2016}, numerical solutions of the transport equation through finite-difference schemes \cite{Corti_2019, bisschoff_2019, PELOSI2025} and solutions through stochastic approaches \cite{Helmod_cuda, Song_2021, guzman_2024, Fiandrini, NT2017}.
}
It is important to note that a detailed and in-depth understanding of the fundamental physical processes requires a physics-based solution, although such models are computationally demanding and often affected by parameter degeneracies and difficulties in constraining all the inputs. This challenge is increasingly being addressed through modern computational techniques and hardware improvements \cite{COSMICA}. However, as highlighted in several reviews \cite{Liu_2024, Whitman2019}, physics-informed models with many parameters do not necessarily produce more accurate results than simplified models, and may in fact be less suitable for forecasting applications.
Since the focus of this study is on developing a forecasting tool with a limited set of physically motivated parameters to reduce degeneracy, we adopted a one-dimensional numerical model of GCR transport in the heliosphere.

\REV{This model, result of a collaboration between the \emph{Università degli Studi di Perugia} (Pg) and the \emph{Laboratório de Instrumentação e Física Experimental de Partículas} in Lisbon (Lis), incorporates the fundamental physical processes of particle transport, such as diffusion, advection, and adiabatic cooling, as well as the charge-sign dependence, needed to compute the energy spectrum and temporal evolution of cosmic radiation in the inner heliosphere}. We identified the most suitable set of model parameters for our model resorting to information criteria techniques. The forecasting strategy establishes empirical relationships between model parameters and solar activity proxies, explicitly accounting for the time lag between these quantities. This framework enables the estimation of cosmic-ray fluxes in advance, based solely on the observed state of solar activity.

\subsection{Outline}
This paper is organized as follows. Sect.\,\ref{sec:calc} introduces the physics of GCR propagation in the heliosphere, describing the radial Parker equation together with the numerical scheme employed to solve it. The information-criteria framework used to identify the most appropriate parameter space is then presented in Sect.\,\ref{sec:BIC}. The calibration strategy, based on multi-channel datasets and a global fitting procedure, is outlined in Sect.\,\ref{sec:calib}, where we also illustrate the model's performance in reconstructing GCR spectra at different epochs. A detailed account of how the monthly timeseries of the model parameters are obtained, filtered, and correlated with solar proxies can be found in Sections\,\ref{sec:temporal_evolution}, \ref{sec:EMD}, and \ref{sec:cross_correlation}.
A discussion of the bootstrap-based approach adopted to estimate model uncertainties is provided in Sect.\,\ref{sec:uncertainty}.
In Sect.\,\ref{sec:Results}, we present our cosmic-ray forecasting results, against several validation datasets, using the predictive capabilities enabled by time-lag and sunspot number forecasts.
The model's application to radiation-dose calculations is discussed in Sect.\,\ref{sec:dosimetric}.
Finally, Sect.\,\ref{sec:conclusion} summarizes the main results, highlights current limitations, and outlines possible improvements and future applications of the model.

\section{Methods}
\subsection{The numerical solution}
\label{sec:calc}
The propagation of GCRs can be conceptualized in two main subsequent steps: first, their transport in the Galaxy from the sources to the heliospheric boundary, where, at equilibrium, the flux corresponds to the LIS; and second, the transport from the heliospheric boundary to Earth.
\REV{The latter is described by Parker's transport equation \cite{PARKER19659} which can be rewritten as}:


\begin{equation}
\label{eq:parker}
\begin{split}
    \frac{\partial\psi}{\partial t} &= \nabla\cdot\left(\textbf{K}_s\cdot\nabla\psi - \mathbf{V}_\mathrm{sw} \,\psi + \langle\textbf{V}_{\mathrm{D}}\rangle \,\psi \right) \\
    &+ \frac{1}{3 p^2} (\nabla \cdot  \mathbf{V}_\mathrm{sw}) \frac{\partial (p^3\psi)}{\partial p} + Q
\end{split}
\end{equation}

where $\psi = \psi(\vec{r}, p, t)$ is the phase-space density, and it relates to the isotropic differential flux $J$ in kinetic energy per nucleon $E_k$ through the  relation $J(E_k) = p^{2}\,\psi(p)$, for a given particle species at time $t$, momentum $p$ , and heliospheric position $\vec{r}$.

\REV{The terms in the transport equation correspond to advection driven by solar wind speed $\mathbf{V}_\mathrm{sw}$, diffusion defined by the symmetric part of the diffusion tensor $\mathbf{K}_s$, pitch-angle-averaged drift velocity $\langle\textbf{V}_{D}\rangle$, adiabatic energy losses associated with the divergence of the solar wind, and an additional source term $Q$ \cite{Moraal2013}.}
The core feature of the PgLis model is the identification of the polarity-dependent cross-correlations between the modulation parameters and a chosen solar proxy. To this end, we implement a radial solution of the Parker equation, solved numerically through a Crank-Nicolson implicit scheme, as seen in \citet{PELOSI2025}.

The Parker transport equation can be reduced to a second-order parabolic equation in space and a first-order equation in rigidity (see Eq.\,\ref{eq:Parker_radial}). This form is obtained by assuming spherical symmetry, a radially outward solar wind with constant velocity $V_\mathrm{sw}$, and an isotropic diffusion tensor, together with the steady-state condition ($\partial \psi / \partial t = 0$) and the absence of source terms ($Q = 0$).

\REV{At this stage, the model requires a parameterization of the effective diffusion coefficient $K$, which is directly related to the parallel diffusion coefficient $K_{\parallel}$ within $\mathbf{K_s}$. These parameterizations are typically based on quasi-linear theory (QLT), which has been successful in describing parallel diffusion \REV{\cite{Jokipii_1966}}. Within this framework, parallel diffusion is governed by resonant scattering between particles of rigidity $P$ and magnetic-field fluctuations.
The parallel mean free path $\lambda_{\parallel}$ depends primarily on the power spectrum of the IMF and can be expressed as $\lambda_{\parallel} \sim P^{2} / w(k_{\rm res})$ \cite{Fiandrini}, where $k_{\rm res}$ is the resonant wave number and $w(k_{\rm res})$ is the power of the magnetic-field fluctuations at the resonant scale. If the power spectral density follows a power law of the type $w(k) \sim k^{-\nu}$, with the index $\nu$ determined by the turbulence regime and spatial scales, then the parallel mean free path scales as $\lambda_{\parallel} \sim P^{\,2 - \nu}$ \cite{engelbrecht_2022}.
The parallel diffusion coefficient, related to the mean free path through $K_{\parallel} = \beta c\, \lambda_{\parallel} / 3$, therefore inherits the same dependence as the turbulence spectrum (see \cite{Engelbrecht_review, Fiandrini, Shalchi2004} and references therein).}

\REV{The perpendicular diffusion coefficient is often written in terms of the parallel diffusion coefficient ($K_{\perp} \propto K_{\parallel}$) \cite{Corti_2019, Fiandrini} despite this being an approximation \cite{burger_2000}. In particular, when solving the equation in 1D, these two terms are merged into a single $K$ coefficient which weights their contributions, while keeping the approximation that both follow the same power law.
Other works have included radial and angular dependencies in the diffusion coefficient \cite{Fiandrini, Corti_2019}, usually through their parameterization of the IMF or as multiplicative factors. Given the application of this work, we have sought to estimate a single effective term which averages and encompasses these effects into a $k_0(t)$ coefficient which follows the data in time (as seen in Eqs.\,\ref{eq:linear}, \ref{eq:single}).
}

\REV{In literature, several functional forms have been proposed to describe the rigidity dependence of the diffusion coefficient, following the available observations and QLT predictions \cite{bieber_1994, Shalchi2004, Engelbrecht_review}. These range from linear or single-power-law expressions \cite{NTPRL, BOSCHINI_helmod, bieber_1994, engelbrecht_2022} to more sophisticated double-power-law models featuring a smooth transition around a characteristic rigidity, see Eqs.\,(4, 5) in \citet{Ngobeni_2020} for details. The latter are particularly suited to three-dimensional modulation frameworks, where drift effects, detailed turbulence properties, turbulence-transport coupling, and their impact on the particle mean free path, are explicitly considered \cite{Corti_ICRC2023}.}

To fulfill the goal of building a 1D effective model that accurately describes the GCR spectrum, we chose to test from the following diffusion coefficients, keeping in mind our sensitivity and the degeneracies that may rise from choosing models with more parameters, and their calibration using available data:
\begin{align}
	&K_1(P,t\,;\,k_0) = k_0(t)\, \beta(P)\, \frac{P}{P_0} \label{eq:linear}\\
	&K_2(P,t\,;\,k_0,\delta) = k_0(t)\, \beta(P)\, \left( \frac{P}{P_0} \right)^{\delta(t)} \label{eq:single}
\end{align}
where $P_0 = 1\,\mathrm{GV}$, $\beta = v/c$ embeds the mass/charge dependence, and $k_0$ is a normalization coefficient of order $10^{22}\,\mathrm{cm^{2}\,s^{-1}}$.
As argued in \citet{PELOSI2025}, we also fix the radial solar wind speed and the modulation boundary (heliopause location) to the same values, namely $V_\mathrm{sw} = 450\,\mathrm{km\,s^{-1}}$ and $r_{\mathrm{HP}} = 122\,\mathrm{AU}$.

Charged particles traveling through the heliosphere display organized gradient and curvature drift motions which depend on local magnetic conditions, their charge sign, and the polarity of the solar magnetic field. This effect, together with diffusion and solar-wind advection, constitutes one of the components of solar modulation of GCRs~\cite{Jokipii1977}.

\REV{
As seen in Eq.\,(\ref{eq:parker}), the drift term $\langle\mathbf{V}_{D}\rangle$ enters the transport equation as an advective contribution, with the requirement of being divergence-free. If we want to modify the advection term in order to capture charge-sign dependence and be physically consistent, we need to keep this in mind.
}

\REV{Given the 1D nature of this numerical solution, particle motion is averaged over azimuthal and zenithal directions, meaning that the advective term included in this solution describes the net average effective radial motion of particles.
This effective contribution cannot be described by drift physics directly, its a net effect on the radial advection of particles that can be parameterized.}

\REV{
We thus introduce a radial effective velocity $V_\mathrm{c}(t) = \varepsilon(t)\, V_\mathrm{sw}$, which enters the transport equation as an additional radial advective term alongside $V_\mathrm{sw}$.
Following this logic, the radial transport equation takes the form,
\begin{equation}
    \begin{split}
        &\nabla_r \cdot \left(K\cdot \nabla_r \psi\right) - \left(\nabla_r \cdot V_\mathrm{sw}\right)\,\psi - \left(\nabla_r \cdot V_\mathrm{c}\right)\,\psi \\
        &+ \frac{1}{3\,p^2} \left( \nabla_r \cdot V_\mathrm{sw} \right) \frac{\partial \left(p^3\,\psi\right)}{\partial p} = 0,
    \end{split}
\label{eq:Parker_streaming_radial}
\end{equation}
where $K$ is the effective diffusion coefficient, spatially constant.}

\REV{
In this form of the Parker equation, expanding 
$\nabla_r\cdot(V_\mathrm{sw}\,\psi)$ produces a term proportional to $\psi\,\nabla_r\cdot V_\mathrm{sw}$; however, this same contribution appears with opposite sign in the adiabatic energy-loss term, which is also proportional to $\nabla_r\cdot V_\mathrm{sw}$, and therefore cancels out in the final radial equation. By contrast, since no assumption is made on the divergence of $V_\mathrm{c}$,  expanding $\nabla_r\cdot(V_\mathrm{c}\,\psi)$ produces an analogous term proportional to $\psi\,\nabla_r\cdot V_\mathrm{c}$; this term has no counterpart in the adiabatic loss, which involves only 
$\nabla_r\cdot V_\mathrm{sw}$, and therefore does not cancel, leaving a residual contribution in the radial equation.}

\REV{
In the steady-state approximation, in the absence of sources, and under spherical symmetry, the 1D transport equation can be rewritten as
\begin{equation}
    \begin{split}
        &K\frac{\partial^{2}\psi}{\partial r^{2}} + \left( \frac{2K}{r} 
        - (1+\varepsilon)\,V_\mathrm{sw} \right)\frac{\partial\psi}{\partial r} \\
        &- \frac{2\,\varepsilon\,V_\mathrm{sw}}{r}\psi 
        + \frac{2V_\mathrm{sw}}{3r} \frac{\partial\psi}{\partial\ln p} = 0
    \end{split}
    \label{eq:Parker_radial}
\end{equation}
where the $-2\,\varepsilon \,V_\mathrm{sw}/r\,\psi$ term is the residual contribution from $\nabla\cdot V_\mathrm{c}$, and the effective diffusion coefficient $K$ is radially constant.}


The time-dependence of the modulation parameters will later be shown to be linked to the 11-year solar activity cycle guaranteeing that the temporal evolution of our solution is slow, ensuring that the system can be considered to have reached equilibrium between time intervals, keeping the steady-state approximation valid.
This temporal dependence allows us to capture charge-sign polarity dependence of radial advection. As will be seen later, even though the diffusion coefficient and its power-law index show similar correlation patterns with solar proxies for both polarities, the behavior of $\varepsilon$ highlights the charge-sign polarity coupling.

\begin{figure*}[t]
	\centering
	\includegraphics[width=0.8\textwidth]{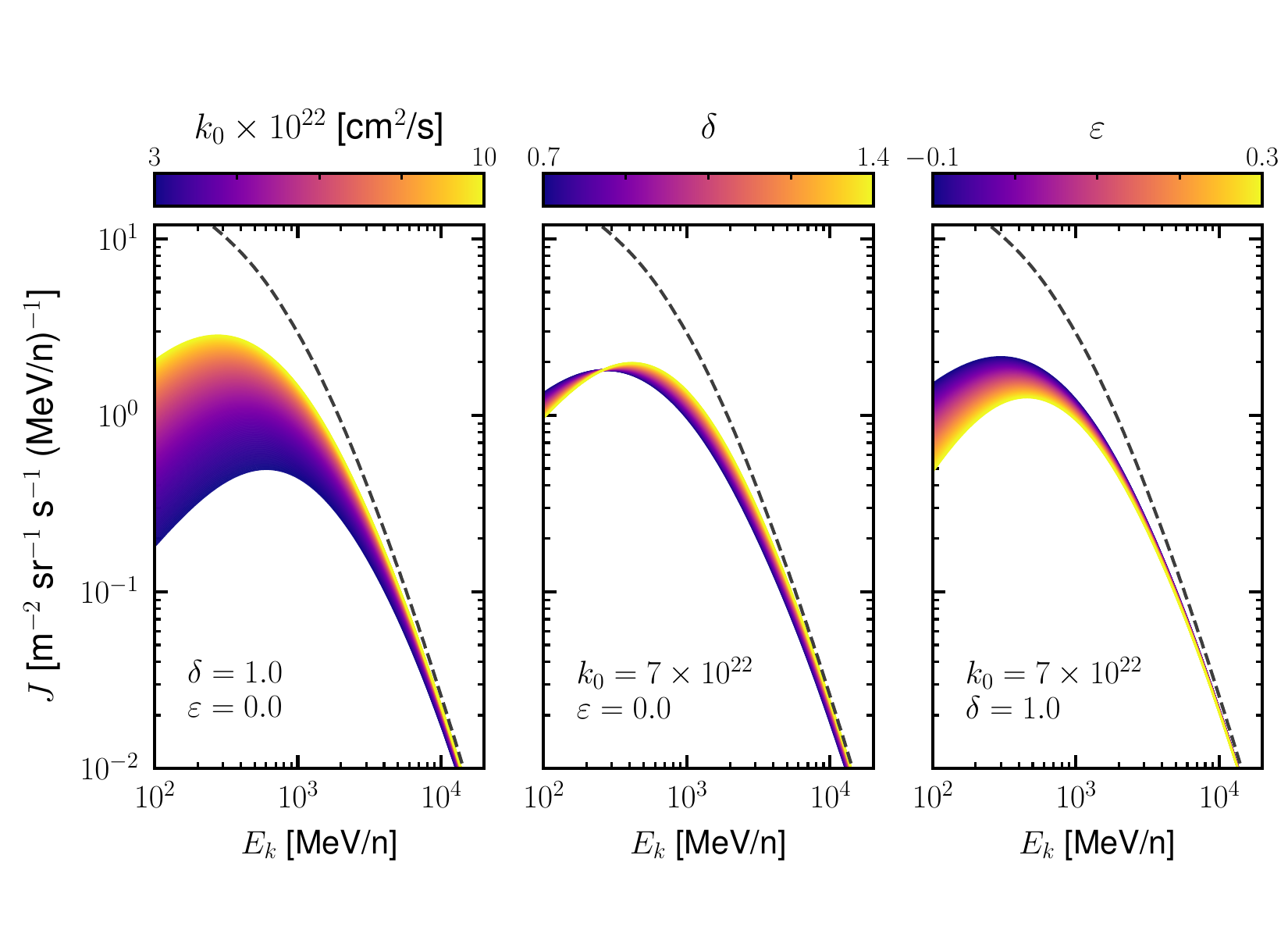}
	\caption{
		Individual effects of propagation parameters on the proton flux near Earth, obtained by varying each parameter independently within the range determined by the fitting procedure, while keeping all others fixed. The gray dashed line indicates the proton LIS used for calibration. See the main text for a detailed discussion of how each parameter affects the modulated flux.
	}
	\label{fig:effects}
\end{figure*}

Figure\,\ref{fig:effects} illustrates how the effective advection parameter shapes the spectrum, showing that it predominantly affects the low-energy region by modifying the curvature of the flux. This is also showing evidence that this term captures the change of diffusive regime that occurs at low rigidities.
It is important to note that despite the addition of this term, the model is not sensitive to other features such as the asymmetric structure of the heliosphere, the effects of the termination shock and heliopause, or the complex geometry of the global IMF \cite{POTGIETER_review}.

To identify the most appropriate parameter space, we adopt an information-criteria approach.

\subsection{The information criteria}
\label{sec:BIC}
Information criteria offer a straightforward way to quantify the information content of a parameter set, particularly in the presence of degeneracies. They provide an objective way to compare different parameterizations by balancing model performance against complexity. A simple comparison based on maximum likelihood is inadequate, since models with more parameters can always fit the data better without necessarily offering improved physical insight or predictive power, while reducing the number of degrees of freedom. Information criteria overcome this limitation by introducing a complexity penalty, thus favoring parameterizations that achieve an optimal balance between accuracy and parsimony. As applied in \cite{Liddle} and \cite{GODLOWSKI}, we adopt the Bayesian Information Criterion (BIC), introduced in \cite{BIC_Schwarz}, defined as
\begin{equation}
	\text{BIC} = -2 \ln(\mathcal{L}) + k_{\tiny{\mathrm{params}}} \ln(N),
\end{equation}
where $\mathcal{L}$ is the maximum likelihood of the model, obtained from the \(\chi^2\) statistic through $\mathcal{L} \propto \exp\left(-\chi^{2}/2\right)$, calculated from Eq. \ref{eq:chi2}, $k_{\tiny{\mathrm{params}}}$ is the number of free parameters, and $N$ is the number of data points used in the fit. The preferred model is the one with the lowest BIC value.

\subsection{Model calibration}
\label{sec:calib}
The calibration of the model parameter set $\vec{q}$ is based on time- and energy-resolved proton measurements from AMS-02 and PAMELA, together with monthly carbon fluxes from ACE. For AMS-02, we use the daily proton fluxes recorded between May 2011 and November 2019 \cite{AMS_PRL2021}, averaged over Bartels rotations (BR)\footnote{Bartels rotation = 27 days, Carrington rotation = 27.27 days}, covering energies from $0.45\,$GeV to $100\,$GeV. For PAMELA, we used the Carrington-rotation-averaged proton fluxes, collected on board the Resurs-DK1 satellite from June 2006 to January 2014 \cite{PAMELA_2013_protons, PAMELA_2018_protons}, spanning energies from  $88\,$MeV to $46.5\,$GeV. Since our model assumes species-independent transport parameters for GCR nuclei, we also include monthly ACE/CRIS carbon observations, which cover an energy range from $\sim\!50$ to $\sim\!200$~MeV/n, thereby exploiting the overlapping energy and time ranges among the three experiments.
Fits are performed independently for each BR using the standard \textsc{MINUIT} routine. For a given BR, the minimized function is
\begin{equation}
	\chi^{2}(\vec{q}) =
	\sum_{s} \sum_{j} \sigma_{s,j}^{-2}
	\left[
	\hat{J}_{s}^{\,j}
	- \frac{1}{\Delta P_j}
	\int_{P_j}^{P_{j+1}}
	J_{s}(P, \vec{q}) \, dP
	\right]^{2},
	\label{eq:chi2}
\end{equation}
where, for each GCR species $s$, the model flux $J_{s}(P,\vec{q})$ is integrated over the experimental rigidity or energy bin $\Delta P_j$, and $\hat{J}_{s}^{\,j}$ is the corresponding measured flux. The terms $\sigma_{s,j}$ represent the experimental uncertainties.

\begin{figure*}[t]
	\vspace{0.4cm}
	\centering
	\includegraphics[width=0.8\textwidth]{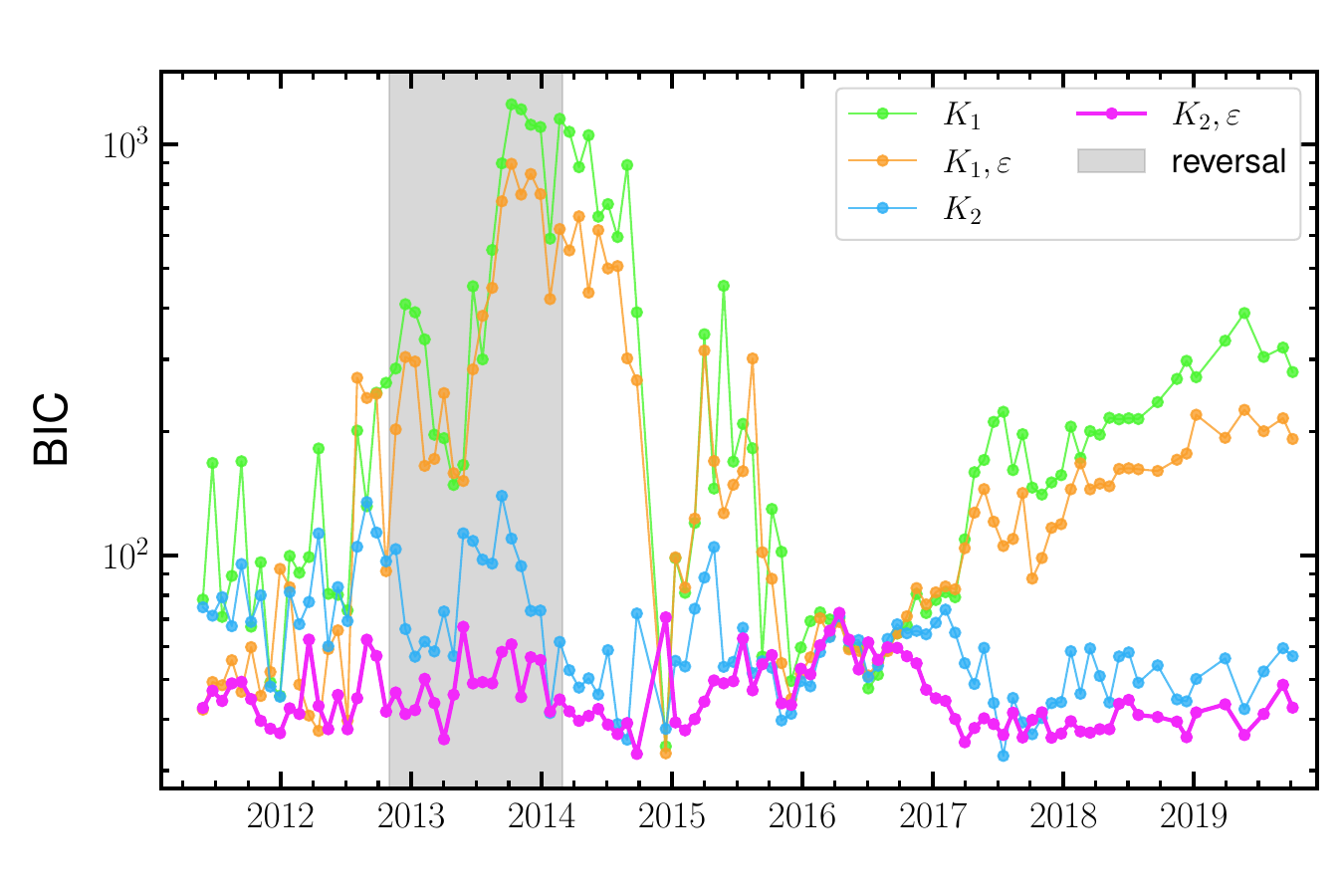}
	\caption{
		Bayesian Information Criterion (BIC) values obtained from fits to AMS-02 proton data and ACE/CRIS carbon data, for the different combinations of model parameters listed in Sect.\,\ref{sec:calc}. The parameterization $K_2$ (Eq.\,\ref{eq:single}), when coupled with $\varepsilon$, consistently yields the lowest BIC values, indicating a significant statistical preference. Its stable behavior over the entire time range, including the polarity-reversal phase (gray band), shows that this parameterization reproduces the data robustly while reducing parameter degeneracy.
	}
	\label{fig:BIC}
\end{figure*}

For the parameter space selection we rely on AMS-02 proton data, which provides the highest statistics and the smallest uncertainties, complemented by ACE carbon measurements. For both proton and carbon LIS, we adopt the calculations of \cite{LIS_Boschini_2020}.
Figure \ref{fig:BIC} shows the BIC values obtained for the three diffusion-coefficient parameterizations. Following the conventional interpretation \cite{jeffreys1961theory}, a BIC difference of $2$ already constitutes positive evidence, while a difference of $6$ or more provides strong evidence against the model with the larger BIC.
Our results clearly indicate that the parameterization $K_2$ (Eq.\,\ref{eq:single}) coupled with $\varepsilon$ yields the smallest BIC values, significantly lower than those obtained with the alternative formulations. Moreover, its behavior remains stable over time, without abrupt variations even during the polarity-reversal period, when solar modulation effects are strongest. This stability shows that the parameterization is capable of reproducing the observed flux shapes consistently across different solar phases and modulation conditions.

The free parameters to be fitted are therefore $\vec{q} = \{k_0, \delta, \varepsilon\}$.
To illustrate the individual effect of each modulation parameter on the modulated flux near-Earth, we vary one parameter at a time while keeping the others fixed, as shown in Fig.\,\ref{fig:effects}. It is evident that, despite correlations among parameters, each of them influences the flux shape in a distinct way.

\REV{The normalization parameter $k_0$ reshapes the flux, shifting the peak toward lower energies as its value increases.
It affects the solution in the whole energy range, up to $10^4$ MeV, and is often the most impactful parameter used to describe solar modulation.
Plenty of solutions rely solely on this parameter to describe this phenomenon, as is the case of~\citet{PELOSI2025} and other 1D solutions, the BON model~\cite{BON2020}  and even the force-field approximation~\cite{Moraal2013}.
The parameter $\delta$ shifts the spectrum horizontally, moving the peak toward higher energies for larger values of $\delta$. As seen in Sec.~\ref{sec:calc}, this parameter relates to the frequency spectrum of the turbulence of the heliospheric magnetic field. It is allowed to vary with time in order to capture local magnetic conditions as they evolve with the solar activity cycle.
The parameter $\varepsilon$ mostly regulates the low-energy curvature of the spectrum, in a shorter energy range than $k_0$ (up to $10^3$ MeV); in particular, lower values of $\varepsilon$ correspond to a stronger flattening of the flux at low energies, allowing for better agreement with very low-energy experimental data such as that measured by ACE.
}

\begin{figure*}[!t]
	\centering
	\includegraphics[width=0.8\textwidth]{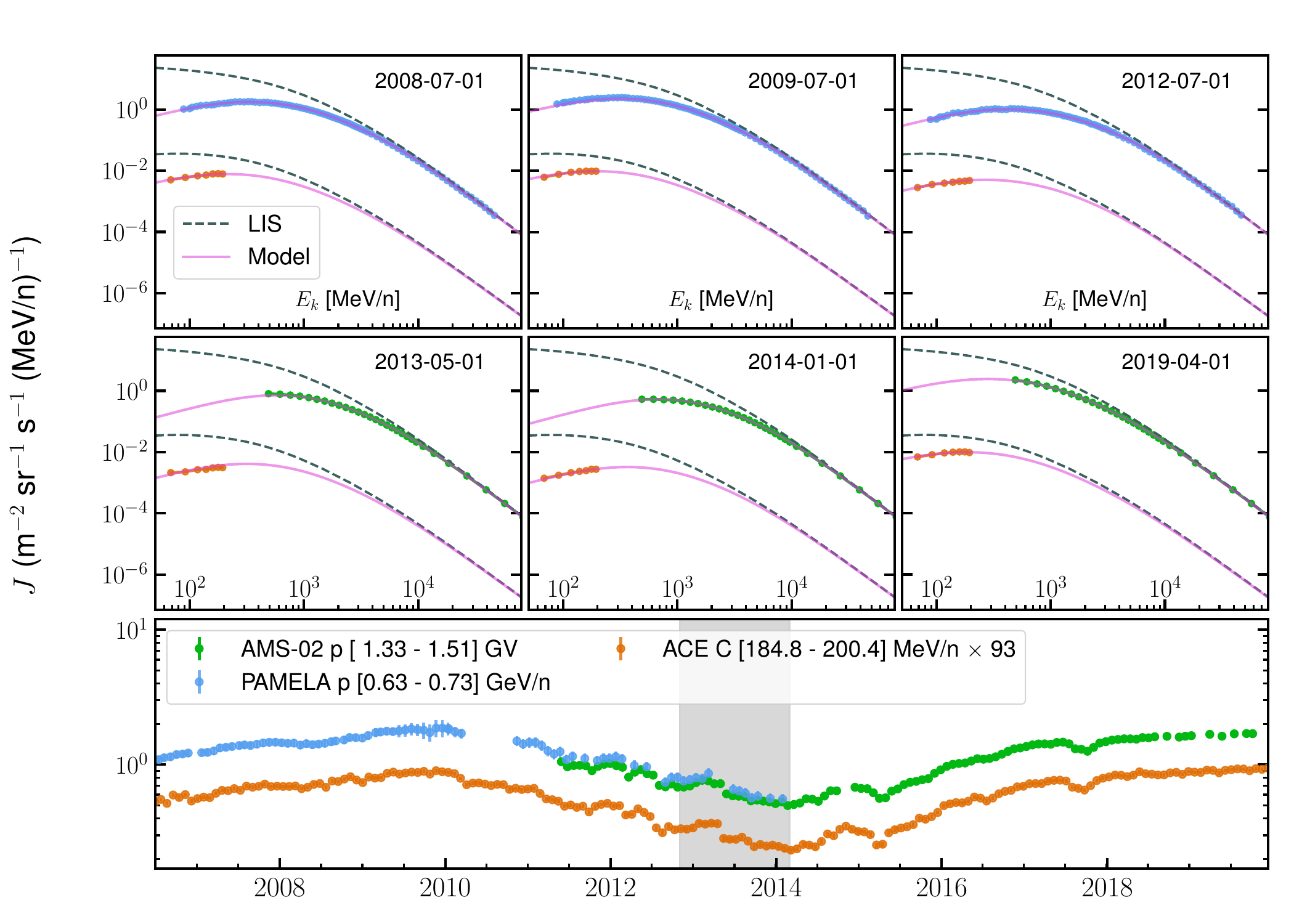}
	\caption{
		Modulated energy spectra for protons measured by PAMELA (light blue) and AMS-02 (green), and for carbon measured by ACE (orange), shown for six selected BRs. The magenta lines indicate the best-fit fluxes, while the gray dashed lines represent the local interstellar spectra (LIS) for protons and carbon adopted in this work \cite{LIS_Boschini_2020} as inputs to the model. The model shows good agreement with the data across all epochs considered. The bottom panel displays the temporal evolution of the proton and carbon fluxes measured by PAMELA, AMS-02, and ACE (scaled for visualization purposes) at selected reference energy bins. The gray shaded band marks the polarity-reversal phase, following the description of \citet{Sun_reversal_2013}.
	}
	\label{fig:model_performance}
\end{figure*}

As discussed in Sec.\,\ref{sec:calc}, we adopt a quasi-steady approach in which each monthly dataset is matched to a steady-state solution of the transport equation with its corresponding parameter set $\vec{q}$. This approximation is justified by the characteristic timescales of solar-modulation variability being longer than, or at least comparable to, the monthly resolution of the experimental datasets used in this work \cite{NTGalaxy}.
A crucial benefit of using calibration data from 2006 to 2019 is that this interval spans a wide range of solar conditions across two solar cycles (23 and 24), which is essential for the correlation study at the core of the PgLis model. It includes the deep solar minimum during the negative-polarity epoch ($\text{A}<0$; 2006–2009), the subsequent rise toward the solar maximum around early 2014, and the following descending phase in the positive-polarity epoch ($\text{A}>0$).
The performance of the model for six selected BRs is shown in Fig.\,\ref{fig:model_performance}. The magenta lines represent the best-fit proton and carbon fluxes, while the gray dashed line indicates the LIS used in the calculations. The fits were performed independently for each BR using the minimization procedure defined in Eq.\,\ref{eq:chi2}.
The results show that the model successfully reproduces the cosmic-ray flux across all phases of solar activity and for both species considered, over the full extended energy range.

\subsection{Temporal evolution of the modulation parameters}
\label{sec:temporal_evolution}
\begin{figure*}[t]
	\vspace{0.4cm}
	\centering
	\includegraphics[width=0.8\textwidth]{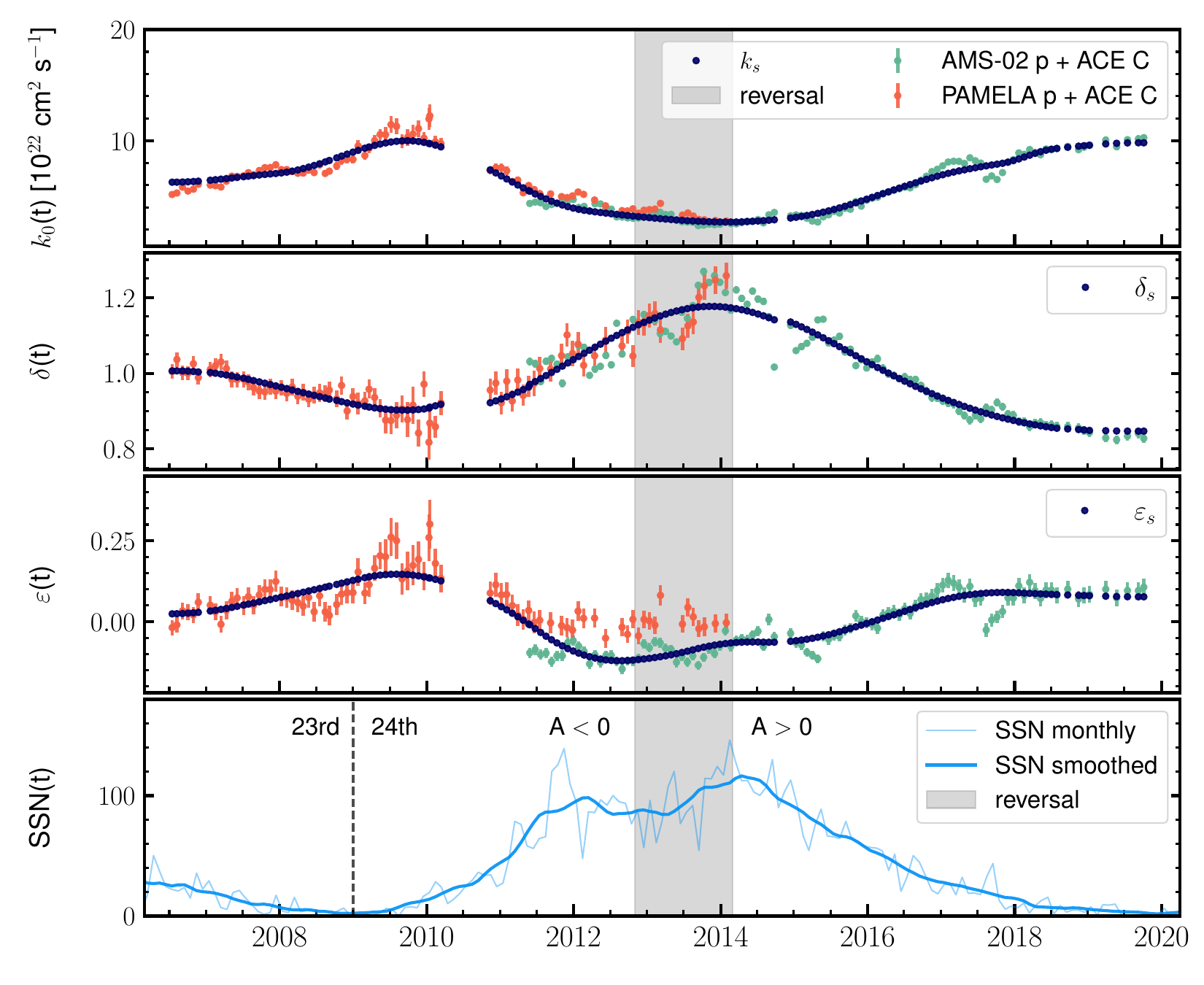}
	\caption{
		Best-fit results for the model parameters $k_0$, $\delta$, and $\varepsilon$ obtained from the fitting procedure described in Sec.~\ref{sec:calib}, using proton measurements from PAMELA and AMS--02 combined with monthly carbon data from ACE. The dark blue points show the same parameters after smoothing with the HHT filtering algorithm described in Sec.~\ref{sec:EMD}. The bottom panel displays the temporal evolution of the monthly and smoothed sunspot number (SSN), highlighting its correlation with the modulation parameters. The shaded band marks the polarity-reversal phase, following the description of \citet{Sun_reversal_2013}.
	}
	\label{fig:timeseries}
\end{figure*}

The fitting procedure was applied to all epochs corresponding to the PAMELA and AMS-02 measurements, combined with monthly ACE carbon data, producing timeseries of best-fit modulation parameters $\vec{q}$, shown in Fig.\,\ref{fig:timeseries}.

The bottom panel displays the monthly sunspot number (SSN) together with its 13-month smoothed version, as provided by the SILSO/SIDC center at the \textit{Royal Observatory of Belgium} (Brussels, Belgium)\footnote{Data taken from https://sidc.be/SILSO/home.}.
The propagation parameters exhibit a clear temporal evolution that traces the overall pattern of solar activity, paving the way for the cross-correlation study described in Sect.\,\ref{sec:cross_correlation}.
In addition, the parameters show evident mutual correlations, which will be taken into account in the estimation of the model's uncertainty (see Sect.\,\ref{sec:uncertainty}).
The slight bias observed between the parameter time series derived from the AMS-02 and PAMELA datasets (both complemented with ACE data), particularly the discrepancy in $\varepsilon$ near the polarity-reversal phase, originates from a mild tension between the fluxes measured by the two experiments. This tension is associated to differences in the respective analysis methodologies.
As discussed in \cite{PAMELA_2018_protons}, the proton fluxes are provided as Carrington-averaged values, in contrast to the Bartels-rotation-resolved AMS-02 data. Moreover, days affected by Forbush decreases were removed from the PAMELA dataset, leading to a modest overestimation of the monthly averaged flux during periods of high solar activity.
These effects are likely responsible for the systematic bias observed in the fitted parameter timeseries over the overlapping time range. For this reason, when applying the smoothing algorithm described in Sec.\,\ref{sec:EMD}, we rely only on AMS-02 data during the overlapping period. These measurements provide a more robust temporal reference, as no Forbush-day removal was applied and only rigidity bins contaminated by solar energetic particles were excluded.

\subsection{The smoothing algorithm}
\label{sec:EMD}
As extensively discussed in our previous work \cite{PELOSI2025}, GCR fluxes exhibit both recurrent and non-recurrent short-term variations. These include the well-known 27-day rotational modulation and shorter 9-13.5 day periodicities associated with solar-wind structures \cite{AMS_PRL2021}, as well as stochastic disturbances caused by solar flares or interplanetary coronal mass ejections. Such effects, including Forbush decreases \cite{Aslam_Forbush}, can significantly perturb the daily flux, particularly in the lower-energy range. Since the goal of this work is to model and forecast the long-term evolution of the GCR flux, and to cross-correlate it with a solar proxy (namely SSN), it is necessary to suppress short-term fluctuations to avoid overfitting. For this reason, we apply a smoothing filter based on the Hilbert-Huang Transform (HHT).
The HHT consists of two steps. The first is the Empirical Mode Decomposition (EMD) \cite{EMD}, an adaptive and data-driven method that decomposes a signal into a finite set of Intrinsic Mode Functions (IMFs) $c_k(t)$ and a residual trend $r(t)$:
\begin{equation}
	y(t) = \sum_{k=1}^{N_\mathrm{IMF}} c_k(t) + r(t).
\end{equation}
Each $c_k(t)$ ideally represents an oscillatory mode with a locally well-defined frequency. EMD is particularly effective for non-linear and non-stationary timeseries, and it is widely used in disciplines such as climatology and solar-physics timeseries analysis \cite{Sun2015,Reda}.
In the second step, the Hilbert Transform is applied to each IMF:
\begin{equation}
	\hat{c}_k(t) = \frac{1}{\pi}\,\mathcal{P}
	\int_0^{\infty} \frac{c_k(t')}{t - t'}\,dt',
\end{equation}
where $\mathcal{P}$ denotes the Cauchy principal value.
This yields the analytic signal:
\begin{equation}
	\label{analytic}
	\zeta_k(t) = c_k(t) + i\,\hat{c}_k(t)
	= a_k(t)\,e^{i\varphi_k(t)},
\end{equation}
with instantaneous amplitude $a_k(t)$ and phase $\varphi_k(t)$, from which the instantaneous frequency $\omega_k(t) = d\varphi_k(t)/dt$ is obtained. 
This formulation is particularly well suited for IMFs, or for ensembles of IMFs, due to their amplitude and frequency characteristics. In particular, the estimation of instantaneous frequency remains meaningful as long as the signal satisfies the narrow-band condition \cite{boashash1992}.

\begin{figure*}[!t]
	\centering
	\includegraphics[width=0.8
		\textwidth]{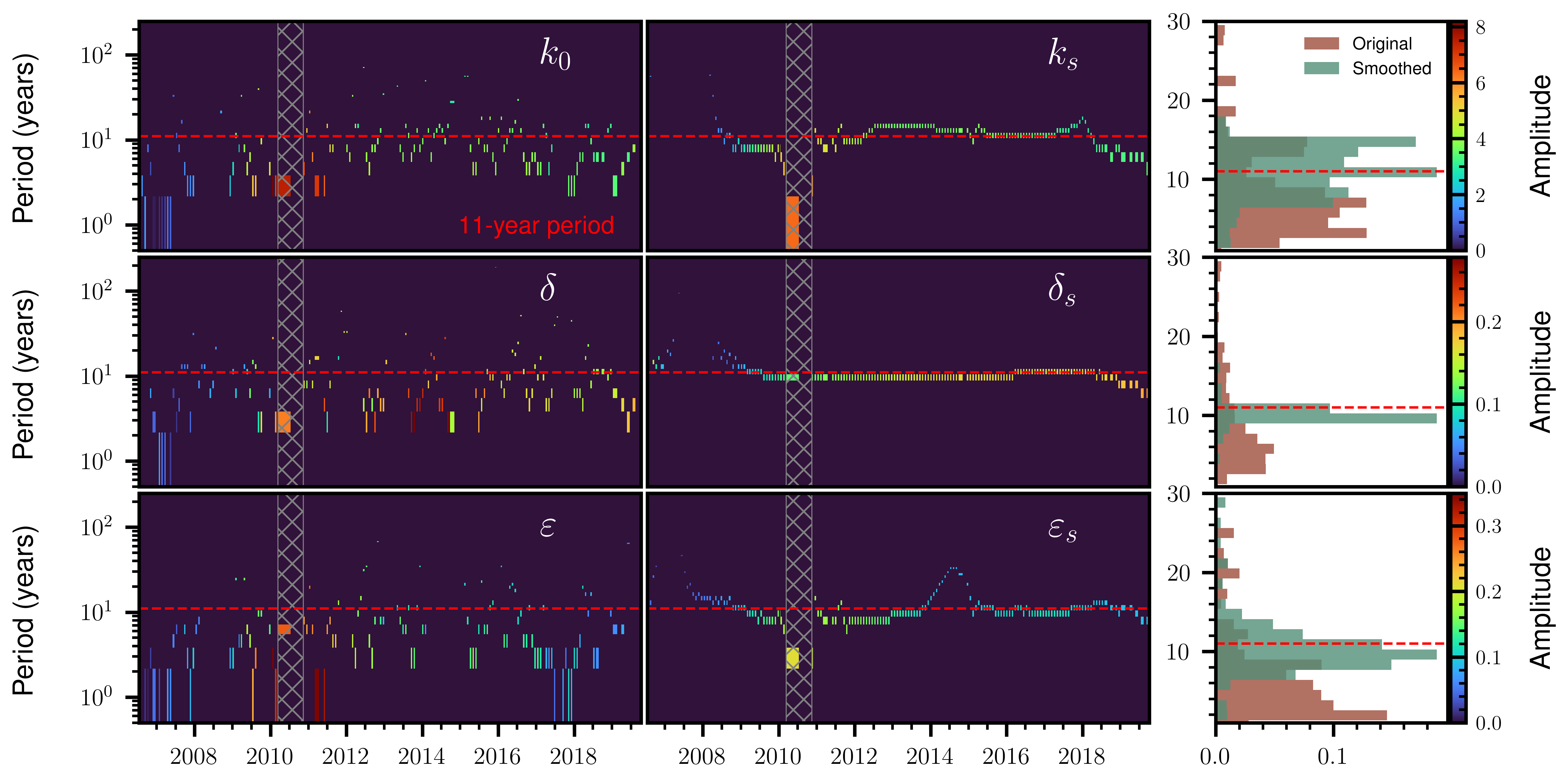}
	\caption{
		Hilbert spectra of the model parameters, showing the frequency content of both the original time series $\vec{q}$ (left) and their smoothed reconstructions $\vec{q}_s$ (right), obtained using the HHT-based algorithm described in Sec.\,\ref{sec:EMD}. For both cases, the long-term trend has been removed. The hatched region indicates the data gap in the PAMELA dataset. The smoothing procedure effectively suppresses high-frequency fluctuations, isolating the long-term trends associated with solar-cycle variability of $\simeq 11$\,years, indicated by the dashed red line. On the right side of each panel, the corresponding frequency content and spread of each spectrum are reported.
	}
	\label{fig:hht}
\end{figure*}

Differently from our previous work, where the optimal set of IMFs was selected by maximizing the Spearman rank correlation between the reconstructed signal and the SSN proxy, here we focus exclusively on the time series of the model parameters and on their frequency content.
Guided by the current understanding of space-climate periodicities and characteristic timescales, we define a loss function that encodes the key properties that a smoothed, long-term time series must satisfy.
First, the reconstructed series must retain good agreement with the original data, as quantified by the $\chi^{2}$ statistic.
Second, its dominant mean frequency should remain close to the typical solar-cycle scale,  $\mu_{\odot} \simeq 11$\,yr, which characterizes long-term heliospheric variability. Third, the spectral content of the reconstruction should exhibit limited dispersion around the dominant frequency.
These requirements are combined into the loss function $\mathcal{C}$ as follows:
\begin{equation}
	\label{eq:cost_function}
	\mathcal{C} =
	w_{\chi}\, Z(\chi^{2})
	+ w_{\mu}\, Z\!\left(|\mu - \mu_{\odot}|\right)
	+ w_{\sigma}\, Z(\sigma),
\end{equation}
where $w_{\chi}$, $w_{\mu}$, and $w_{\sigma}$ are the weights assigned to each component, $\mu$ and $\sigma$ denote respectively the mean frequency and its standard deviation obtained from the Hilbert spectrum of each reconstruction realization, and $Z(\cdot)$ is a normalization operator ensuring that all three contributions enter $\mathcal{C}$ with comparable scale.
Because the EMD depends on several internal parameters, including sifting stopping criteria, relative thresholds, the number of mirror points used to handle boundary effects, the maximum number of iterations per IMF extraction, and the extrema detection algorithm, we perform a refined grid search over the main EMD parameters. For each configuration, we additionally scan over all possible combinations of IMFs allowed by that EMD setup, in order to identify the reconstruction that minimizes $\mathcal{C}$ as defined in Eq.~\eqref{eq:cost_function}.
The Hilbert spectra of each model parameter, for both the original time series and their smoothed reconstructions obtained through the procedure described above, are shown in Fig.\,\ref{fig:hht}, clearly illustrating the reduction of high-frequency content in the smoothed time series. The temporal evolution of the smoothed parameter set $\vec{q}_s$, compared with the original parameters, is also presented in Fig.\,\ref{fig:timeseries}.

\subsection{Cross-correlations}
\label{sec:cross_correlation}
\begin{figure*}[t]
	\vspace{1em}
	\centering
	\includegraphics[width=0.8\textwidth]{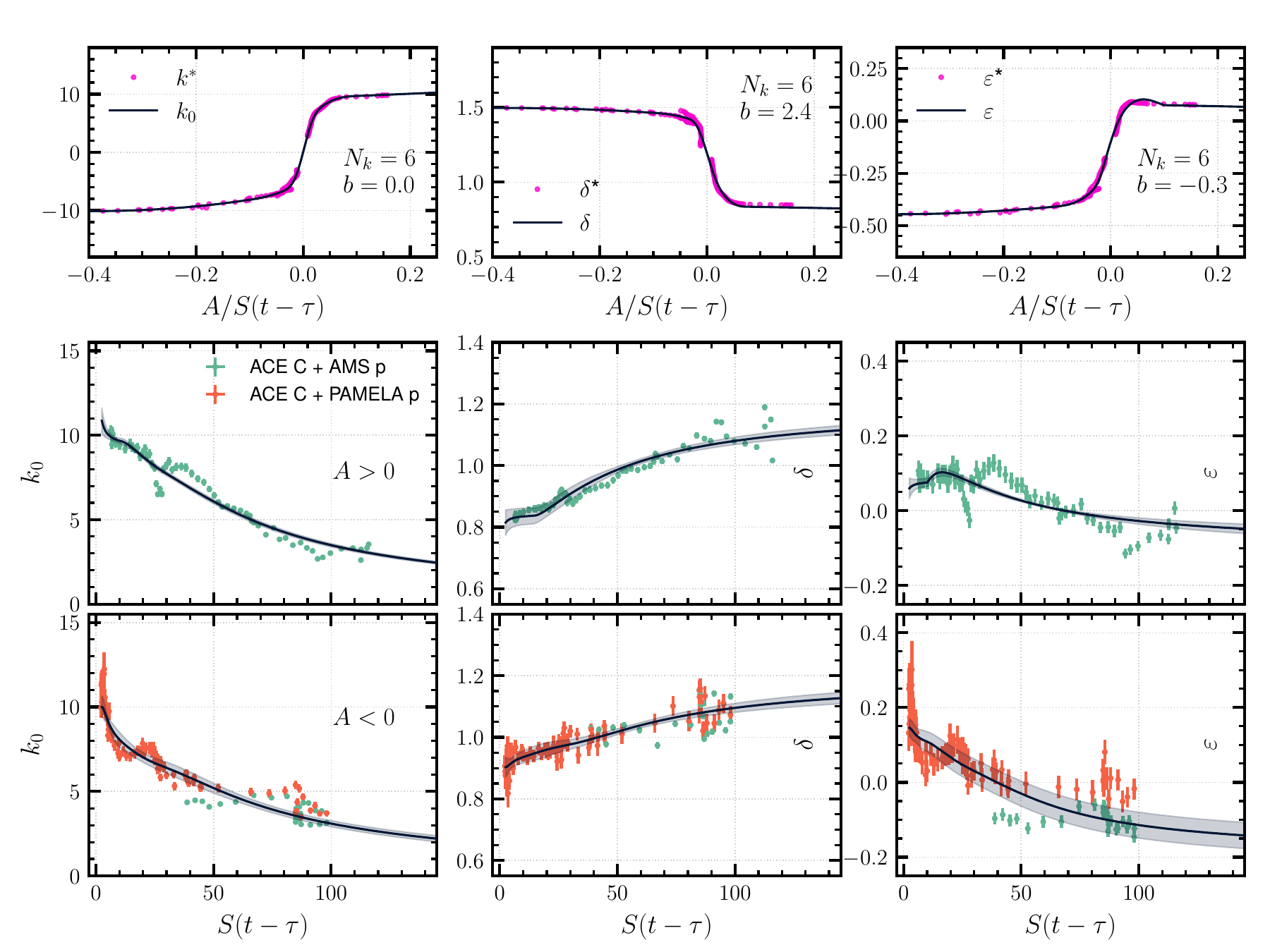}
	\caption{
		\textbf{Top row}: Cross-correlation functions $\vec{f}$ for each model parameter, obtained by minimizing the loss function defined in Eq.\,\ref{eq:loss_score}.
		The magenta points correspond to the transformed parameters $\vec{q}^{\,*}$ defined in Eq.\,\ref{eqnarray:bias_trasform}, plotted against the proxy $\mathcal{\text{A}}/S(t-\tau)$. The fitted values of $N_{\text{knots}}$ and $b$ for each parameter are indicated in the respective panels. \textbf{Bottom rows}: The same functions $\vec{f}$ are plotted against $S(t-\tau)$ for the negative and positive polarity phases, and are shown together with the original, non-smoothed parameter values $\vec{q}$ for reference. These results demonstrate that the long-term evolution of the model parameters is effectively captured by the derived cross-correlation functions. The gray shaded band indicates the 68\% C.L. uncertainty band of each function, obtained using the bootstrap technique described in Sec.\,\ref{sec:uncertainty}.
	}
	\label{fig:splines_global}
\end{figure*}

The core of the PgLis model is the establishment of correlation functions between the smoothed modulation parameters $\vec{q}_s(t)$ and solar proxies, enabling the prediction of GCR fluxes based on solar-activity indicators. As extensively discussed in \citet{Wang_Chi2} and \citet{Fiandrini}, several challenges arise when attempting to establish such correlations: the need to separate the analysis by polarity phases, where distinct patterns are evident; the need to exclude the reversal phase, in which the relation between parameters and proxies is no longer one-to-one; and the need to account for time lags between solar activity proxies and the GCR response in order to remove hysteresis effects. In this study we retain $S$, a continuous time-dependent function representing the smoothed SSN as the model's input solar proxy, and we adopt the time-lag formulation of \citet{tomassetti}, described by the empirical relation:
\begin{equation}
	\label{eq:lag}
	\tau(t) = \tau_{M} + \tau_{A} \cdot \cos\left[\frac{2\pi}{T_{0}}(t-t_{p})\right],
\end{equation}
with parameter values given in Table~2 of \cite{PELOSI2025}.
We also follow the same strategy as in \cite{PELOSI2025} to mitigate the limited coverage of large SSN values, particularly in the anomalously weak Solar Cycle 24, exploiting the effective nature of the model. The approach is based on a redefinition of the proxy as $\text{A}/S(t-\tau)$, where $\text{A}=\pm1$ denotes the IMF polarity.
This transformation ensures a smooth transition during polarity reversal, when $S$ approaches its maximum, and allows continuity to be imposed on the model parameters by using a single global spline fit across both polarities.
The transformed parameters are defined as
\begin{eqnarray}
	\label{eqnarray:bias_trasform}
	q^{*}_{+}(t) & = & q_{s,+}(t) \cdot \text{A}, \\
	q^{*}_{-}(t) & = & b + \text{A} \cdot q_{s,-}(t), \nonumber
\end{eqnarray}
where $b$ is an offset chosen to ensure a smooth connection between the positive and negative branches.
We adopt the same quantile-based knot-placement strategy used in \cite{PELOSI2025} to set the number and positions of the knots $N_{\text{knots}}$. For any given parameter, and for each selected pair $(N_{\text{knots}}, b)$, the correlation function $f$ is modeled as a linear combination of cubic ($k=3$) B-spline basis functions $B_j$:
\begin{equation}
	\label{eq:spline_basis}
	f(x) = \sum_{j=0}^{\,N-k-2} c_j\, B_j(x),
\end{equation}
where $N$ is the number of data points and the coefficients $c_j$ are determined through a
least-squares fit.
The optimal correlation function is obtained by minimizing the loss function
\begin{equation}
	\label{eq:loss_score}
	\mathcal{S} =
	\mathcal{N}\!\left(
	\sum_{i=1}^{N}
	\frac{(y_i - f(x_i))}{|f(x_i)|}^{2}
	\right)
	+
	\mathcal{N}\!\left(
	\int f''(x)^{2}\, dx
	\right),
\end{equation}
where $x_i$ denotes the value of the transformed proxy, $y_i$ the corresponding value of $q^{\,*}$, and $\mathcal{N}(\cdot)$ the normalization operator. The first term controls the agreement between the data and the spline fit, while the second term penalizes large curvature, enforcing smoothness.
The results of the spline fits applied to $\vec{q}^{\,*}$ as functions of $A/S$ are shown in the top row of Fig.\,\ref{fig:splines_global}, where the spline functions are determined. The bottom row shows the same spline functions expressed in terms of $S(t-\tau)$ for both polarity phases.
Edge effects are controlled by removing the last three months of data at both ends of the time series and performing a linear extrapolation. To obtain the parameter values during the reversal interval, we employ the transition functions defined in Eqs.\,(20) and (21) of \cite{Fiandrini}, which combine the values inferred for the two polarity phases.

\subsection{Model's uncertainty}
\label{sec:uncertainty}
\begin{figure*}[t]
	\centering
	\includegraphics[width=0.8\textwidth]{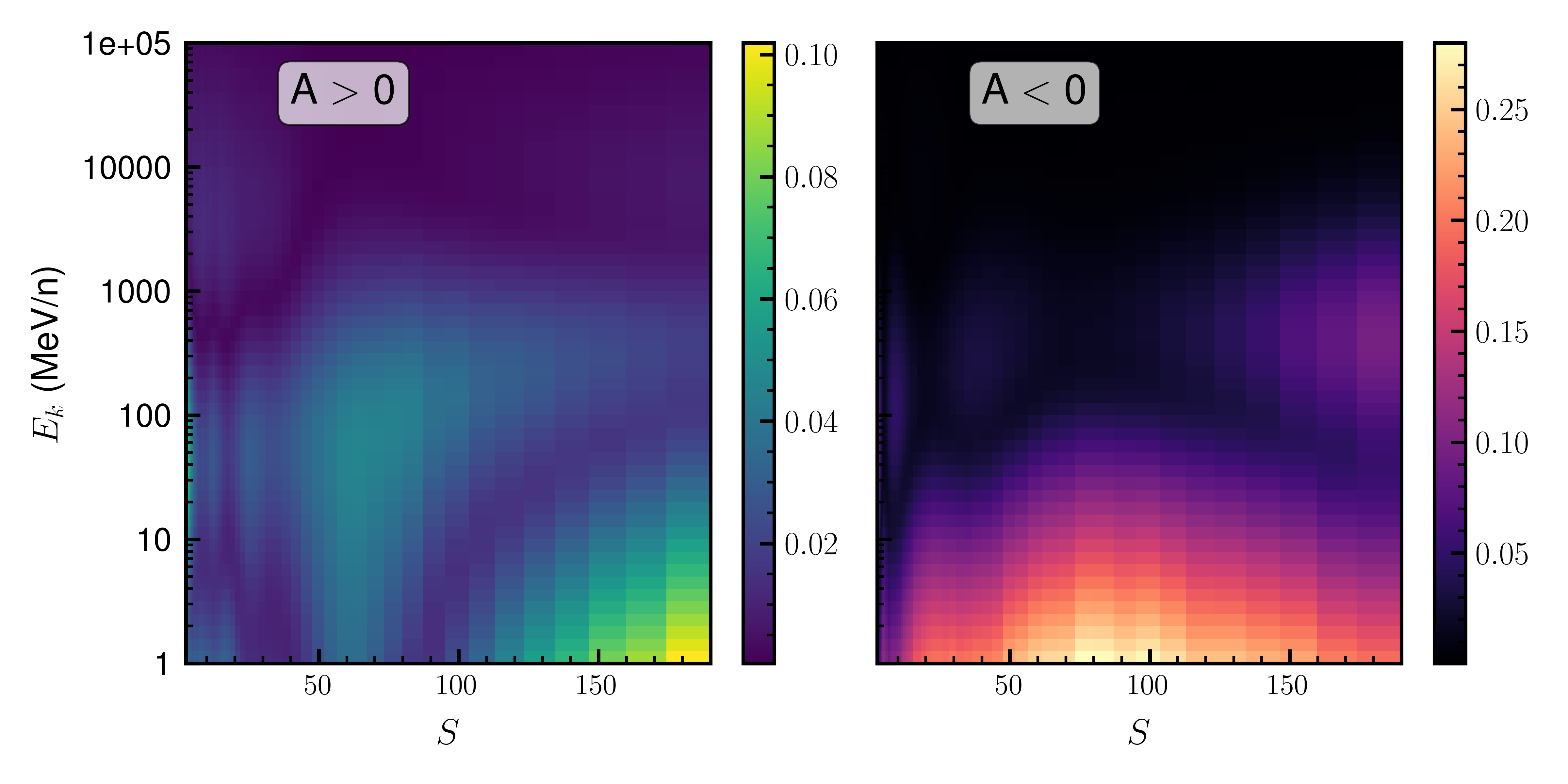}
	\caption{Relative flux reconstruction uncertainty of the PgLis model for protons, computed as a function of the smoothed solar proxy $S$ and kinetic energy $E_k$ for both polarity phases, obtained using the bootstrap procedure described in Sec.\,\ref{sec:uncertainty}.
	}
	\label{fig:bootstrap_uncertainty}
\end{figure*}

Since the HHT-based filtering procedure and the spline-fitting method used in the analysis chain do not natively provide uncertainty estimates, we employ a bootstrap technique to propagate the original uncertainties of the modulation parameters, derived from the fitting procedure, into the final model outputs.
To this end, for each epoch we generate a synthetic parameter vector by sampling from a multivariate Gaussian distribution defined by the best-fit parameter values and the corresponding covariance matrix.
Repeating this process over the full time range yields a simulated time series for each modulation parameter.
We then apply the full HHT-based filtering and spline-fitting procedure described earlier to the ensemble of simulated $\vec{q}$, obtaining a corresponding set of correlation functions for each parameter and polarity phase. The ensemble of these realizations defines the uncertainty of the spline fit, which is taken as the standard deviation of the resulting distribution, as illustrated in Fig.\,\ref{fig:splines_global}.
To propagate these uncertainties to the final flux prediction for a given GCR species, we apply a similar bootstrap strategy starting from a given value of the input solar proxy $S$.
For each value of $S$, we generate a set of model parameter vectors by sampling from a multivariate distribution defined by the fit covariance matrix and the newly derived parameter uncertainties.
By computing the flux for each realization, we obtain an ensemble of spectra, from which the model uncertainty band at any given energy and polarity epoch is determined as the standard deviation of the distribution.
Figure\,\ref{fig:bootstrap_uncertainty} shows the resulting relative uncertainties as a function of the smoothed sunspot number and kinetic energy, for both polarity epochs.
As expected, the uncertainty decreases with increasing energy, reflecting the reduced impact of solar modulation at higher energies. We also observe that the uncertainty in the negative polarity phase is larger than in the positive phase.
This difference directly reflects the weaker constraints on the modulation parameters, which affect the bootstrap procedure, during the $\text{A}<0$ epoch and are driven by the larger uncertainties of the PAMELA measurements compared to the AMS-02 data used during the $\text{A}>0$ phase.

\section{Results}
\label{sec:Results}
\begin{figure*}[t]
	\centering
	\includegraphics[width=0.8\textwidth,trim={0 15px 0 0}]{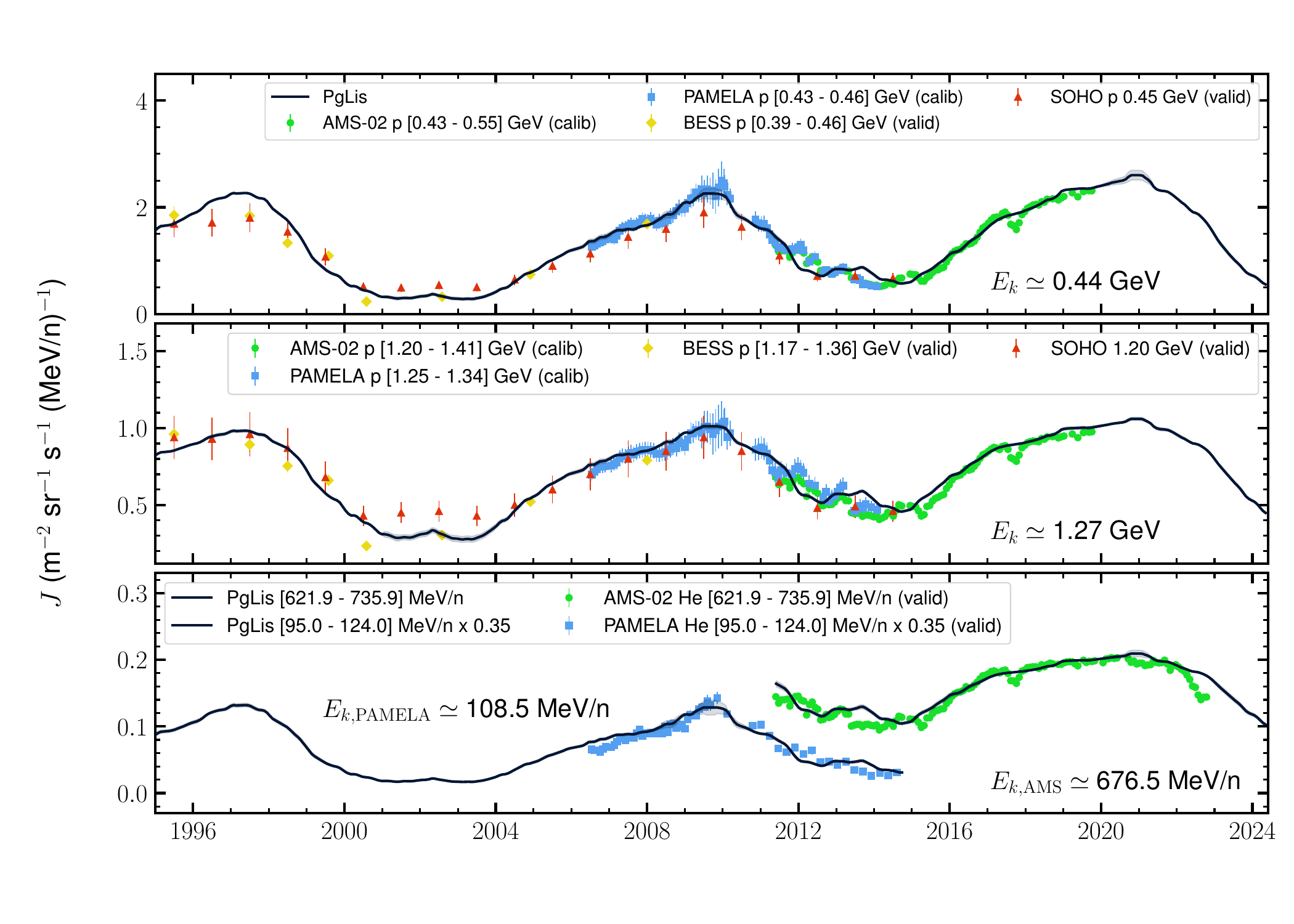}
	\caption{ \textbf{Top}: PgLis predictions spanning several solar cycles at $\sim 440\,\text{MeV}$ for protons, compared with long-term measurements from SOHO/EPHIN and BESS as validation datasets, and with PAMELA \cite{PAMELA_2013_protons,PAMELA_2018_protons} and AMS--02 \cite{AMS_PRL2021} data used for calibration. \textbf{Middle}: Same long-term prediction for multichannel proton fluxes at $\sim 1.3\,\text{GeV}$. \textbf{Bottom}: PgLis predictions compared with helium fluxes from PAMELA \cite{PAMELA_2020_Helium,PAMELA_2022_Helium} (multiplied by a factor of 0.35 for visual clarity) and AMS-02 \cite{AMS_nuclei}, which serve as independent validation datasets. Shaded bands represent the 68\% C.L. model uncertainty. The average relative reconstruction errors for each validation dataset are reported in Table\,\ref{tab:PgLis_uncertainties}.}
	\label{fig:long_term}
\end{figure*}

Given the correlation functions established between the modulation parameters and the solar proxy, together with their associated uncertainties, forecasting the GCR flux at 1\,AU becomes straightforward once the solar input proxy $S$ and the corresponding magnetic polarity $A$ are specified.
At any given epoch $t$, the time-lag relation allows the corresponding model parameters to be evaluated at $t+\tau(t)$.
Solving the Parker transport equation as described in Sec.\,\ref{sec:calc} provides the modulated flux at all radial nodes of the Crank–Nicolson scheme, yielding a reliable solution in particular at $r = 1\,\mathrm{AU}$, where the calibration data lie. This holds for any nuclear species for which a corresponding LIS is available. The calculation assumes universality of the modulation parameters, i.e., that the heliospheric transport coefficients depend only on rigidity. Species-dependent effects arise at the production and galactic-propagation stages, where they shape the LIS, and are further incorporated through the factor $\beta(P)$ in the transport equation, which depends on the particle's mass-to-charge ratio.
To evaluate the long-term accuracy and robustness of the model, we performed a series of performance tests spanning multiple solar cycles, comparing the model predictions with independent experimental datasets. In Fig.\,\ref{fig:long_term} (top and middle rows), we show proton fluxes at approximately $440\,\text{MeV/n}$ and $1.2\,\text{GeV/n}$
from several multichannel experiments, including SOHO/EPHIN \cite{SOHO} and BESS balloon flights \cite{BESS97-2007,BESS93-97}, which serve as independent validation datasets, together with PAMELA and AMS-02 data, which were used for calibration. The figure demonstrates that the model successfully reproduces the long-term evolution of the proton flux over nearly three solar cycles, including the two magnetic polarity reversals that occurred between September 2000 and May 2001 and between November 2012 and March 2014.
The bottom row of Fig.~\ref{fig:long_term} shows helium fluxes at two different energy bins from AMS--02 \cite{AMS_nuclei} and PAMELA \cite{PAMELA_2020_Helium,PAMELA_2022_Helium}.
These data were not included in the calibration and therefore provide an additional, species-level validation, showing good agreement over the entire time span considered.

\begin{figure*}[t]
	\centering
	\includegraphics[width=0.8\textwidth]{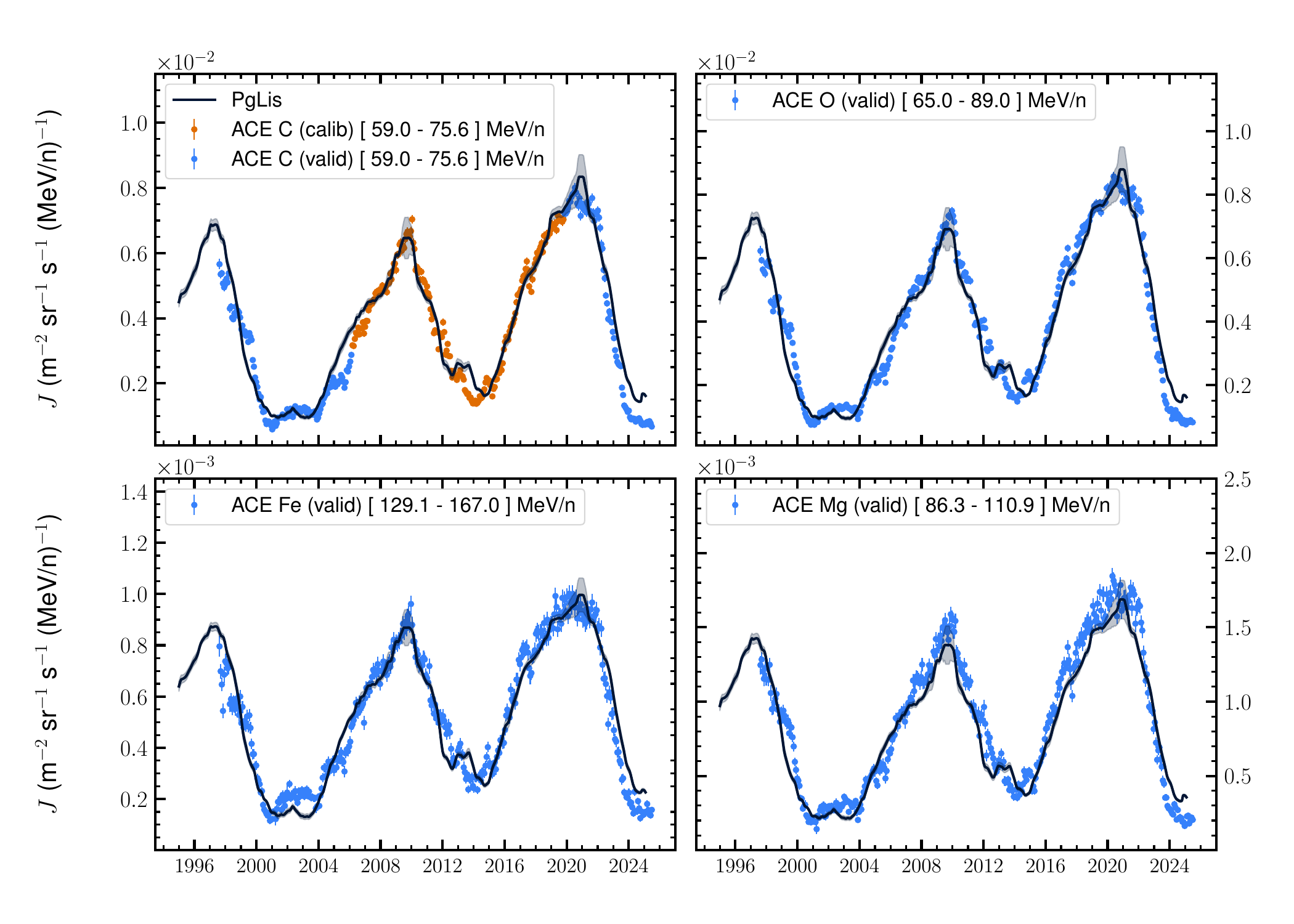}
	\caption{
		PgLis predictions compared with monthly ACE fluxes for carbon, oxygen, iron, and magnesium, spanning several solar cycles. The orange data points indicate the portion of the carbon dataset used for calibration, while the blue points represent the validation datasets. The shaded bands represent the 68\% C.L. model uncertainty. The average relative reconstruction errors for each validation dataset are reported in Table~\ref{tab:PgLis_uncertainties}.
	}
	\label{fig:ACE}
\end{figure*}

To further test the robustness of the model, especially at lower energies, we used monthly flux measurements from ACE/CRIS for several nuclear species.
Figure\,\ref{fig:ACE} shows results for Carbon and Oxygen at energies below 100\,MeV/n, highlighting the portion of the dataset used for calibration. Additional results for Magnesium and Iron, an element of particular importance for dose-evaluation studies \cite{NARICI2012,Naito}, are also displayed. The model reproduces the long-term evolution of all species over multiple solar cycles, demonstrating its universality and stability even in the low-energy regime.

\begin{figure*}[t]
	\centering
	\includegraphics[width=0.8\textwidth,trim={0 15px 0 0}]{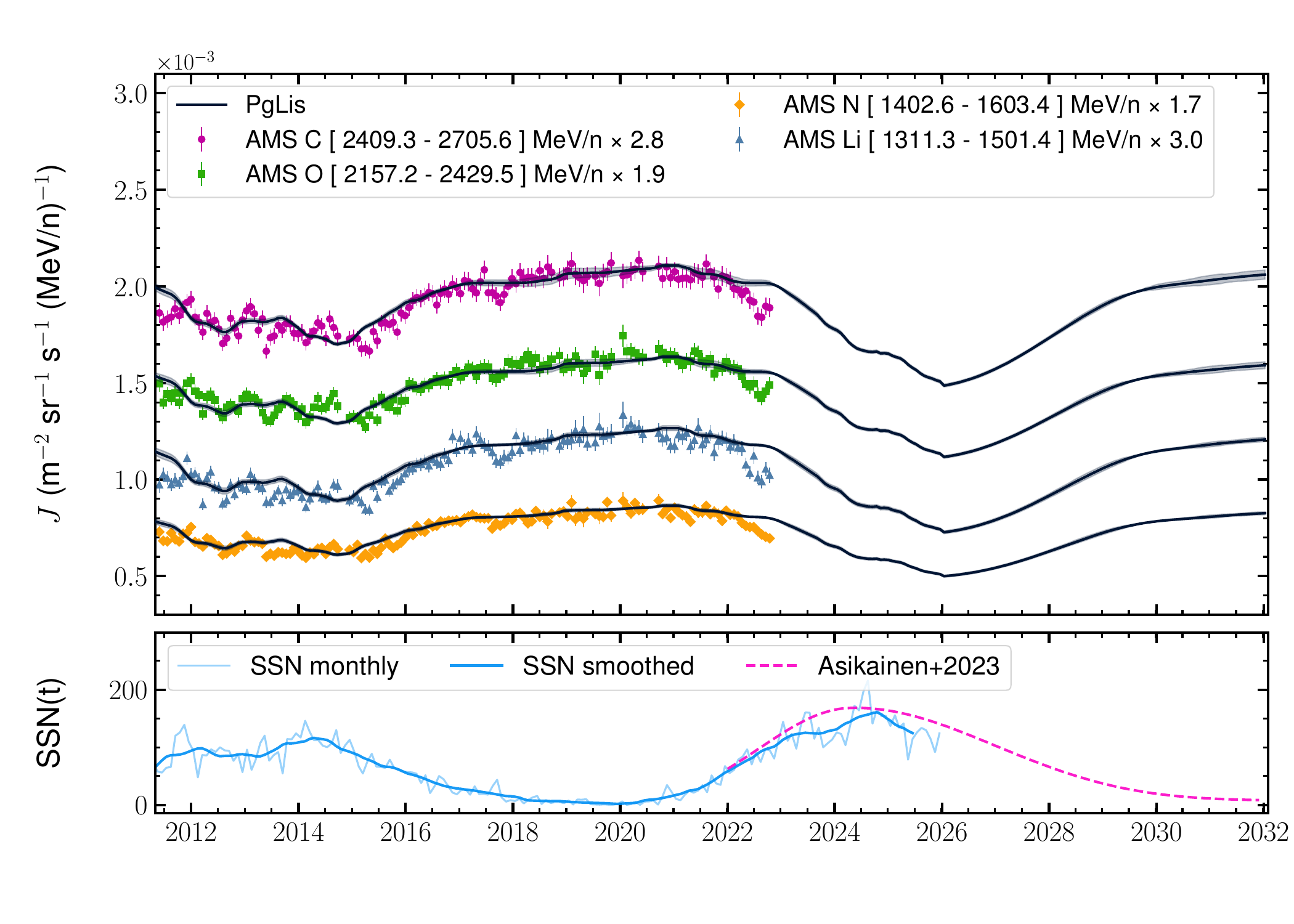}
	\caption{ PgLis predictions compared with monthly AMS-02 \cite{AMS_nuclei} fluxes for carbon, oxygen, nitrogen, and lithium, shown at selected energy bins, from May 2011 to November 2022. These measurements serve as validation datasets. The model, coupled with the SSN forecasts from \cite{Asikainen2023}, provides a prediction for the full 25th solar cycle. The shaded bands represent the 68\% C.L. model uncertainty. The average relative reconstruction errors for each validation dataset are reported in Table~\ref{tab:PgLis_uncertainties}.
	}
	\label{fig:AMS_nuclei}
\end{figure*}

We also tested the model using recent AMS-02 measurements of several nuclear species \cite{AMS_nuclei}, including Carbon, Oxygen, Nitrogen, and Lithium across different energy bins, which serve as an additional validation dataset. The comparison, shown in Fig.\,\ref{fig:AMS_nuclei}, confirms the validity of the universality assumption showing good agreement over the entire AMS-02 operational period from May 2011 to November 2022.
To extend the forecasting capability of PgLis, it is straightforward to couple the model with SSN forecasting models, such as those from the Space Weather Prediction Center of the National Oceanic and Atmospheric Administration (NOAA), as well as predictions from \cite{SSN_future_Penza_2021} and \cite{McIntosh}. In this work, we adopt the forecast of \cite{Asikainen2023} as a reference, shown in the bottom panel of Fig.,\ref{fig:AMS_nuclei}. Using their predicted smoothed SSN as input to the PgLis model, we obtain the flux forecast for the entire Solar Cycle 25. We note that uncertainties in the input SSN prediction should be considered as an additional source of uncertainty in the forecast, together with those arising from the time-lag model and from the adopted LIS within the PgLis framework.
These contributions, which contribute additional uncertainty to the final GCR flux predictions, are not included in the present analysis, where the reported uncertainties are solely those propagated from the forecasting algorithm itself.

Finally, we computed the absolute relative error at each epoch $t$ between the model predictions $f_{\mathrm{model}}$ and the measured fluxes $f_{\mathrm{meas}}$ for all validation datasets, defined as
	\begin{equation}
		\langle R_D \rangle =
		\sqrt{
			\frac{1}{N}
			\sum_{i=1}^{N}
			\left[
				\frac{f_{\mathrm{model}}(t_i) - f_{\mathrm{meas}}(t_i)}
				{f_{\mathrm{meas}}(t_i)}
				\right]^2
		} ,
		\label{eq:error_rms}
	\end{equation}

We then evaluated the average relative reconstruction error for each validation dataset presented in this work from the standard deviation of the corresponding $R_D$ distribution.
The results are summarized in Table\,\ref{tab:PgLis_uncertainties}. The new parameterization, combined with the updated transport-equation solution modeling and the revised frequency-based smoothing procedure, leads to an overall improvement in accuracy across all species and energy ranges tested with respect to the previous model of \citet{PELOSI2025}.

\begin{table}[h!]
	\centering
	\begin{tabular}{l c c c}
		\hline
		Species & $E_k$               & Pelosi+2025           & PgLis                 \\
		        &                     & $\langle R_D \rangle$ & $\langle R_D \rangle$ \\
		\hline
		H       & 95.0 - 124.0 MeV/n  & $18.0\%$              & $10.9\%$              \\
		He      & 621.9 - 735.9 MeV/n & $10.4\%$              & $7.0\%$               \\
		Li      & 1.3 - 1.5 GeV/n     & $6.4\%$               & $4.0\%$               \\
		N       & 1.4 - 1.6 GeV/n     & $6.2\%$               & $4.4\%$               \\
		C       & 59.0 - 75.6 MeV/n   & $24.0\%$              & $18.0\%$              \\
		C       & 2.4 - 2.7 GeV/n     & $3.7\%$               & $2.4\%$               \\
		O       & 65.0 - 89.0 MeV/n   & $24.0\%$              & $16.0\%$              \\
		O       & 2.2 - 2.4 GeV/n     & $4.7\%$               & $2.6\%$               \\
		Mg      & 86.3 - 110.9 MeV/n  & $20.0\%$              & $17.0\%$              \\
		Fe      & 129.1 - 167.0 MeV/n & $20.0\%$              & $16.0\%$              \\

		\hline
	\end{tabular}

	\caption{
		Comparison of the average relative reconstruction error, defined in Eq.~\ref{eq:error_rms}, across all datasets, between the model of \citet{PELOSI2025} and the PgLis model. The PgLis model shows an overall reduction in reconstruction error in every scenario considered.
	}
	\label{tab:PgLis_uncertainties}
\end{table}

\subsection{Dose calculation with PgLis}
\label{sec:dosimetric}
An important application of GCR flux forecasting is the assessment of radiation doses absorbed by astronauts during space missions. Since one of the main goals of this work is to provide a forecasting tool for radiation protection applications, we present here a dosimetric demonstration of its potential.

We consider the scenario of an extravehicular activity (EVA) for an astronaut orbiting Earth on the International Space Station (ISS) and compute the temporal evolution of the body dose due to GCR exposure only.
The first step of this computation is the determination of the position-dependent geomagnetic rigidity cutoff, i.e. the minimum rigidity a particle must have in order to reach a given position within Earth's magnetic field. Assuming quiet geomagnetic conditions not affected by severe solar storms, under which a full computation of the magnetosphere through back-tracing propagation tools would otherwise be required \cite{Larsen_OTSO}, we adopt Størmer's dipolar formulation \cite{smart_1994} for the cutoff to estimate it at any location:
\begin{equation}
    \begin{split}
        P_\mathrm{cutoff} (t, \vec{r}, \varepsilon, \zeta) &= \frac{M(t) \,\mu_0}{4 \pi \, r_\mathrm{dip}^2} \\
        &\times \frac{\cos^4{\lambda(\theta_\mathrm{dip})}}{{\left[1+\sqrt{1-\sin{\varepsilon} \, \sin{\zeta} \, \cos^3{\lambda(\theta_\mathrm{dip})}}\right]}^2},
    \end{split}
\end{equation}
where \(M(t)\) is the time-dependent dipole amplitude which can be derived from the IGRF-13 coefficients \cite{fraser-smith_1987}, \(\mu_0\) is the permeability in vacuum, \(r_\mathrm{dip}\) is the distance to the center of the dipole, \(\lambda(\theta_\mathrm{dip})\) is the co-latitude in the dipole's frame of reference, \(\varepsilon\) is the angle between the particle's direction and the radial direction of the point being evaluated in the dipole's frame of reference, and \(\zeta\) is the angle between the projections of the dipole vector and the particle's direction in the plane perpendicular to the radial direction of the point being evaluated.

Since the cutoff depends on the particle's direction, we assume an isotropic distribution of cosmic rays above the cutoff and estimate a transfer function that assigns a fractional exposure time to each rigidity at any given position inside Earth's magnetosphere.
The ISS orbits Earth with a period of approximately 92 minutes, at an altitude of about 420\,km and with an inclination of roughly 52 degrees. As a consequence, the station samples a wide range of geomagnetic locations during each orbit, leading to time-dependent variations in the effective rigidity cutoff experienced along its trajectory.

The second step consists of the dose calculation. To estimate the total radiation dose experienced by astronauts, it is necessary to determine the average energy deposited by particles of a given energy as they traverse the human body, leading to a complex radiation transport problem.
For practical exposure scenarios, the International Commission on Radiological Protection (ICRP) provides fluence-to-dose conversion coefficients that allow the determination of the average dose received by a human body exposed to an isotropic flux of particles of a given species and energy \cite{ICRP123}. 
The dose can be weighted to account for its biological effectiveness through radiation quality factors, leading to the effective dose $H$, expressed in Sievert (Sv). In this work we adopt the radiation quality factors provided by \citet{icrp_60}, considering whole-body averaged factors that represent the sensitivity of the human body to isotropic radiation.
Therefore, the dose-rate (in Sv/year) calculation can be written as:

\begin{equation}
\label{eq:dose}
\REV{
    \begin{aligned}
        &H(t) =  \sum_{\mathrm{Z}=1}^{28} \;  \Omega_{\mathrm{LOS}} \int_{0}^{\infty} Q_\mathrm{Z}(E_k) \, \frac{D_{\mathrm{Z}}}{\Phi_{\mathrm{Z}}}(E_k) \, J^\mathrm{LEO}_Z(E_k, t) \, \dd E_k, \\
        &H^\mathrm{Shield}(t) = \sum_{\mathrm{Z}=1}^{28} \;  \Omega_{\mathrm{LOS}} \int_{0}^{\infty} Q_\mathrm{Z}(E_k) \, \frac{D_{\mathrm{Z}}}{\Phi_{\mathrm{Z}}}(E_k) \,
        J^\mathrm{Shield}_{\mathrm{Z}}(E_k,t) \, \dd E_k, \\
        & J^\mathrm{Shield}_{\mathrm{Z}}(E_k,t) = J^\mathrm{LEO}_{\mathrm{Z}}(E_k',t)\, \frac{\mathcal{S}(E_k')}{\mathcal{S}(E_k)} \,e^{- \eta \, h}, \\
        &J^\mathrm{LEO}_Z(E_k, t) = \mathcal{P}_\mathrm{Cutoff}(E_k|\vec{r}(t), t) \, J_{\mathrm{Z}}(E_k,t),
    \end{aligned}
}
\end{equation}

\REV{
where $Z$ labels the particle species, \(\Omega_{\mathrm{LOS}}\) represents the allowed solid-angle, \(Q_Z\) is the quality factor for a given nucleus and $D_{\mathrm{Z}}/\Phi_{\mathrm{Z}}$ is the fluence-to-dose conversion coefficient for isotropic radiation. The quantity $J_{\mathrm{Z}}(E,t)$ denotes the differential isotropic GCR flux, computed monthly by the PgLis model. We include all species from protons ($\mathrm{Z}=1$) up to nickel ($\mathrm{Z}=28$).
$J^\mathrm{LEO}_{\mathrm{Z}}(E_k,t)$ represents the flux in low Earth orbit (LEO) and is the solar modulated flux, attenuated by the effect of the geomagnetic field through a BR averaged transfer function $\mathcal{P}_\mathrm{Cutoff}$ which will be described later.
$J^\mathrm{Shield}_{\mathrm{Z}}(E_k,t)$ is the LEO flux after attenuation by a material, equivalent to aluminum, of thickness $h$ in $\mathrm{g}/\mathrm{cm}^2$.
Finally, the attenuation coefficient $\eta = 5 \times 10^{-26} \mathcal{N}_A \left(A^{-1/3} + 27^{1/3} - 0.4)^2\right) / 27$, $\mathcal{N}_A$ is the Avogadro constant, A is the atomic mass of the projectile particle considered, $\mathcal{S}$ is the stopping power of the projectile particle in aluminum, $E_k'$ is the energy of the projectile before crossing the material, and can be written as $E_k' = \mathcal{R}^-1(\mathcal{R}(E_k) + h)$ and $\mathcal{R}$ is the range of the projectile particle in aluminum. This attenuation procedure is described in \citet{Adams1983}.
}

\begin{figure*}[t]
    \centering
    \includegraphics[width=0.8\textwidth]{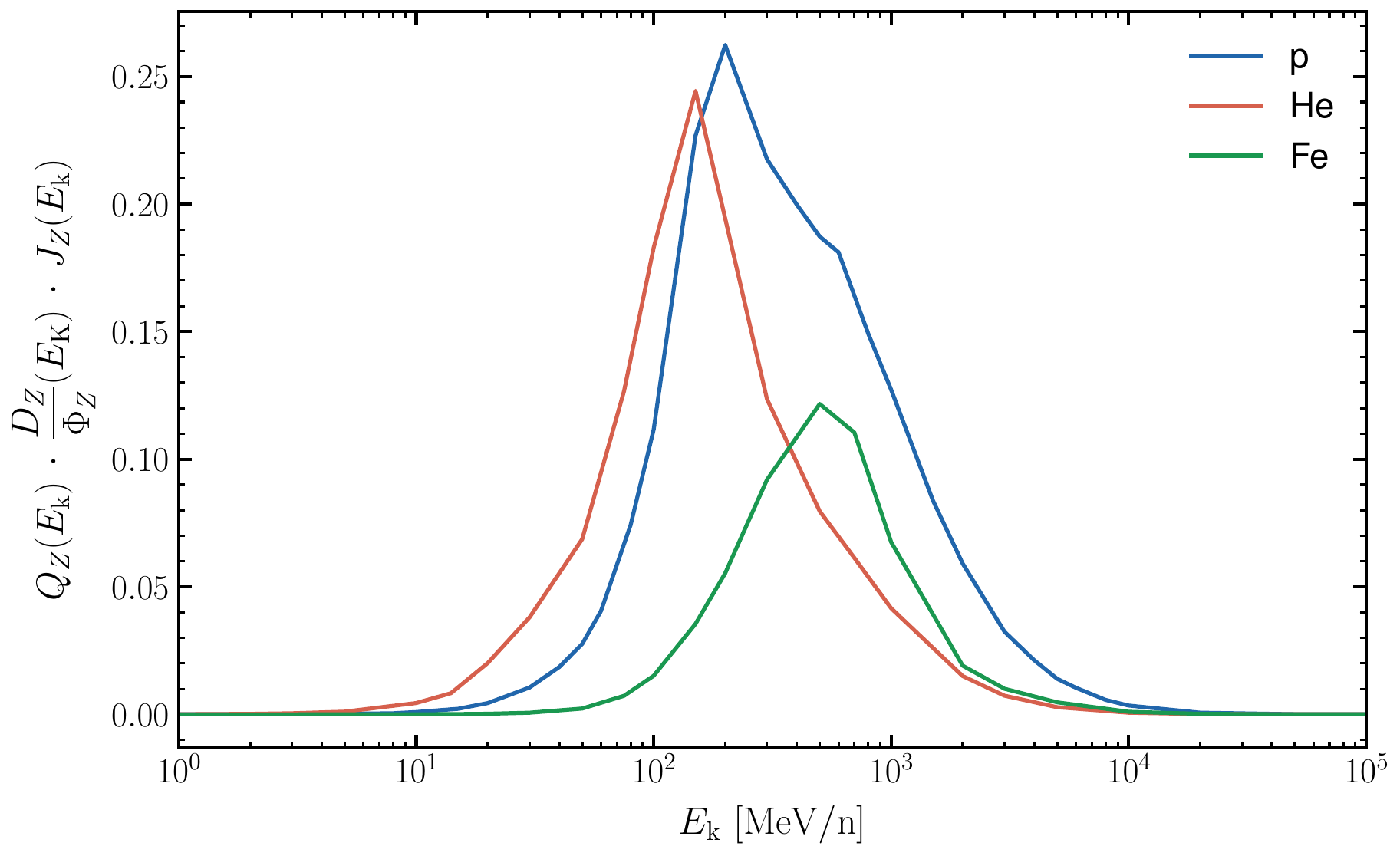}
    \caption{\REV{Integrand $Q_Z(E_k)\cdot D_Z/\Phi_Z(E_k)\cdot J_Z(E_k)$ of the effective dose rate in Eq.\,(\ref{eq:dose}) for protons (blue), helium (red), and iron (green), computed from PgLis fluxes at a reference solar-minimum epoch. The integrand peaks in the range $\sim 100$\,MeV/n to $\sim 1$\,GeV/n, identifying the energy interval most responsible for the radiation dose.}}
    \label{fig:dose_integrand}
\end{figure*}

\REV{Figure\,\ref{fig:dose_integrand} illustrates the energy dependence of the dose-rate integrand $Q_Z(E_k)\cdot D_Z/\Phi_Z(E_k)\cdot J_Z(E_k)$ entering Eq.\,(\ref{eq:dose}), shown for three representative species: protons (blue), helium (red), and iron (green). Despite its relatively low abundance, iron contributes non-negligibly to the total effective dose, highlighting the importance of including all species from $Z=1$ to $Z=28$ for an accurate dose assessment. The integrand peaks in the range $\sim 100$\,MeV/n to $\sim 1$\,GeV/n, identifying the dominant GCR kinetic energy range responsible for the radiation dose.
}


\begin{figure*}[t]
	\centering
	\includegraphics[width=0.8\textwidth]{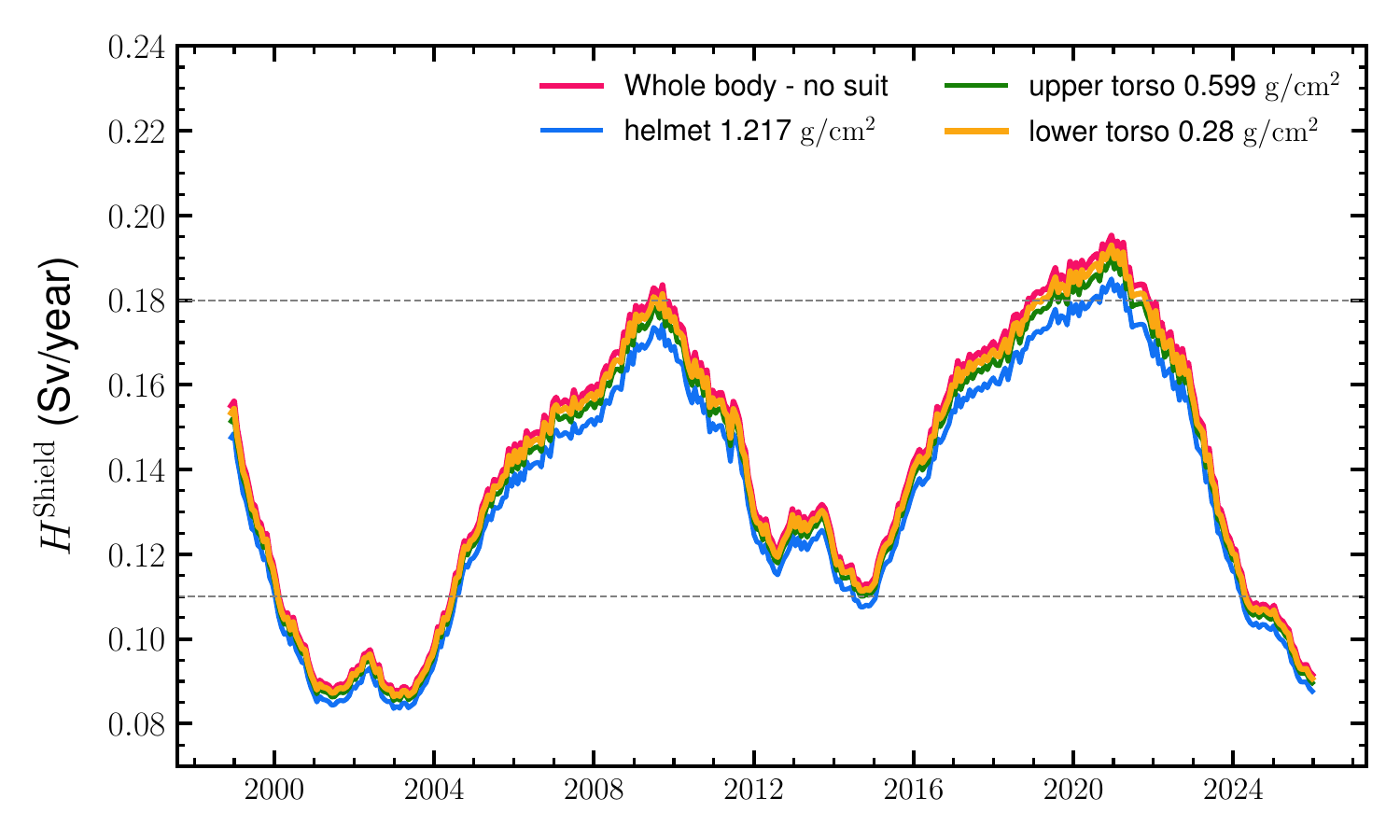}
    
    \caption{Monthly averages of the whole-body effective dose rate (in Sv\,yr\(^{-1}\)), computed using ICRP fluence-to-dose coefficients and quality factors, induced by GCRs (\(\mathrm{Z}=1\)–28) modeled with PgLis for an astronaut in ISS orbit during an EVA. The calculation assumes sex-averaged reference body proportions, accounts for the geomagnetic cutoff using Størmer's dipolar approximation, and considers three shielding configurations corresponding to a representative spacesuit design \cite{Wilson_1997}. The two horizontal gray lines represent the average maximum and minimum effective dose rates absorbed by an astronaut in ISS orbit, as reported in \cite{Restier}. In the legend, the shielding thickness is given as Teflon thickness.}
	\label{fig:body_dose}
\end{figure*}

To account for Earth shadowing of GCR, \(\Omega_{\mathrm{LOS}}\) represents the allowed solid-angle cone at ISS' orbital altitude. Its asymptotic boundary is determined by the zenith angle of the line of sight (LOS) tangent to the Kármán line, conventionally set at an altitude of $100\,\mathrm{km}$ \cite{Adams1991}.
The cutoff is computed as distribution through the transition function $\mathcal{P}_\mathrm{Cutoff}(E|\vec{r}, t)$ averaged per monthly ISS orbital positions \cite{orcinha_2025}.
Given the EVA scenario, we consider several spacesuit shielding configurations, including the helmet, upper torso, and lower torso (arms and legs) of the Shuttle spacesuit, based on the descriptions provided in \cite{Wilson_1997}. For simplicity, Teflon is assumed as a representative material for all components. The shielding is modeled by varying the equivalent aluminum thickness using the thin-wall flux transport model presented in \cite{Adams1983}. The results of this computation are shown in Fig.\,\ref{fig:body_dose}.

\REV{The dose rate $H^\mathrm{Shield}$ computed with PgLis for an astronaut performing an EVA in ISS orbit range from approximately $0.08$ to $0.20$\,Sv\,yr$^{-1}$ over the solar cycles considered (Fig.~\ref{fig:body_dose}), reflecting the anti-correlation with solar activity that characterizes GCR modulation. These values can be contextualized against established radiation-protection limits. Space agencies currently set a universal, age- and sex-independent career limit of $600$\,mSv effective dose for all astronauts \cite{Shavers2024}. At the solar-minimum, dose rates predicted by PgLis ($\sim 0.20$\,Sv\,yr$^{-1}$), indicate that the career limit would be reached in approximately three years of continuous EVA exposure, underscoring the relevance of long-term GCR forecasting for mission planning. At solar maximum, the predicted dose rate drops to $\sim 0.08$\,Sv\,yr$^{-1}$, extending that estimate to roughly seven years. It should be noted that the present calculation accounts for GCR exposure only. During solar energetic particle events (SEP), particle fluxes (mainly protons) at energies of $\sim10$-$100$\,MeV/n can exceed the GCR background by several orders of magnitude; 
however, at these energies particles are effectively attenuated by a few g\,cm$^{-2}$ of shielding material, so that operational countermeasures such as sheltering in more heavily shielded modules can substantially mitigate the acute SEP dose. A quantitative assessment of the SEP contribution is left for future work. For reference, the lethal dose to 50\%  of the human population (LD$_{50}$) is of order $4$\,Gy \cite{Space_rad_lethal}, far above the annual doses discussed here, confirming that the GCR-induced dose in the scenarios considered poses a long-term risk rather than an acute radiation hazard. Reflections such as these are fundamental as we move towards a future in which civil and commercial transportation of humans are more common.}

\section{Discussion and Conclusions}
\label{sec:conclusion}
The evaluation of space-radiation hazards relies on the capability to forecast the radiation environment, in which GCR modulation plays a central role.
Reliable predictions of cosmic-ray fluxes across different nuclear species are essential for planning future space missions and for assessing the radiation risks faced by astronauts and spacecraft systems.

Recent high-precision GCR measurements from PAMELA, AMS-02, ACE, and other missions, together with additional data expected from upcoming space-borne experiments, are playing a key role in improving our understanding of heliospheric modulation and charged-particle propagation.

In this work, we have proposed a new model of GCR modulation in the heliosphere that predicts cosmic-ray fluxes near-Earth from their relationship with solar activity proxy.
We solved a 1D Parker equation using a set of physically motivated modulation parameters, in which the full tensor nature of cosmic-ray diffusion, anisotropic effects, curvature and gradient drifts, and large-scale heliospheric structures are not accounted for.
Including these effects would increase the model complexity and computational cost, while also introducing a large number of additional parameters that are poorly constrained and would need to be known for forecasting applications.

The strength of the present approach lies in its calibration procedure, which combines the energy coverage and temporal information provided by multiple experiments, and in its reliance on a single solar proxy to describe the evolution of the model parameters, namely the smoothed SSN.
Given the phenomenological nature of the model, the time-dependent parameters should be interpreted as quantities that encode heliospheric conditions averaged over the GCR propagation histories, rather than as direct physical observables. Their temporal evolution must therefore be understood as an effective description of the heliosphere's response to solar activity.

Furthermore, our analysis of the correlation patterns between the model parameters and the time-delayed solar proxy reveals distinct behaviors during opposite polarity epochs, thereby exposing the underlying charge-sign dependence.

To extract these long-term trends, we applied a filtering procedure to remove short-term components associated with stochastic solar activity fluctuations, such as CMEs and flare-driven disturbances, which cannot be captured by the SSN or other solar proxies. This was followed by a penalized spline fit, ensuring a smooth representation of the correlation functions.

As demonstrated in the final section, the model achieves strong accuracy in both flux reconstruction and forecasting, despite the inherent approximations and the limited calibration dataset. Comparisons with measurements from multiple experiments across different energy ranges, epochs, and GCR species confirm this performance. The choice of SSN as the proxy is motivated by its ability to incorporate delayed responses through the time-lag formalism, which leads to well-established correlation patterns with the modulation parameters, as well as by the availability of forecasting models that enable decadal-scale predictions.

Looking ahead, the availability of high-energy-resolved data will allow us to investigate whether calibrating the model separately for the ascending and descending phases of the solar cycle can further improve its accuracy.
This approach is motivated by the observed degradation in reconstruction quality at the onset of Solar Cycle 25, during the positive magnetic polarity and ascending phase, as suggested by Fig.\,\ref{fig:AMS_nuclei}.
A phase-dependent calibration may therefore better capture the distinct modulation conditions characterizing different stages of the solar cycle.

We also plan to explore alternative solar proxies that better describe the evolution of the solar activity cycle than the SSN, such as the open solar magnetic flux \cite{open_solar_flux_Frost_2022}. These proxies may reveal new correlation structures and further enhance predictive performance.
Such an extension will require a dedicated characterization of the time lag for each new proxy, a topic that will be investigated in a forthcoming publication.
Long-term GCR reconstructions, together with an assessment of their associated uncertainties, are also of interest for atmospheric and climate studies that require accurate estimates of cosmic-ray-induced atmospheric ionization over decadal timescales \cite{Eyring_CMIP6_2016, Golubenko_2024, Zheng_2023_GEOSChem_Be}.

Finally, we emphasize that our model shows strong potential for applications in space-weather forecasting and radiation protection. In this work, we present a proof-of-concept application of the model to estimate human radiation dose in space, illustrating its capability for dosimetric studies. A comprehensive validation, including a direct comparison between dose-rate predictions derived from the PgLis fluxes and CRaTER measurements \cite{Spence2013}, will be performed in future work. 


\section{Data Availability}
We provide pre-compiled tables of GCR fluxes computed using the PgLis model. The tables are available in \citet{zenodo}. The dataset includes all particle species from protons to nickel (\(Z=1\)–28), for both solar magnetic polarity conditions, and is distributed in CSV format.
The tables contain differential fluxes \(J(t)\) over a rigidity range from \(40\,\mathrm{MV}\) to \(200\,\mathrm{GV}\), parameterized as a function of delayed sunspot numbers \(S(t-\tau)\) (see Eq.\,\ref{eq:lag}). 
\REV{We also release a user-friendly Python package \cite{pglis_py} designed to query the published dataset and retrieve model flux predictions for any requested epoch and species.}

\begin{acknowledgments}
	We acknowledge support from ASI under ASI-UniPG 2019-2-HH.0, ASI-INFN 2019-19 HH.0, its amendment 2021-43 HH.0, the Italian Ministry of University and Research (MUR) through the program “Dipartimenti di Eccellenza 2023-2027” and FCT under grant 2024.00992.CERN, Portugal.
\end{acknowledgments}

\bibliography{refs}

\end{document}